\documentclass[journal]{IEEEtran}
\usepackage{amsmath,amssymb,amsfonts}
\usepackage{bbm}
\usepackage{graphicx}
\usepackage{textcomp}
\usepackage{xcolor}
\usepackage{graphicx}
\usepackage{epstopdf}
\usepackage{lipsum}
\usepackage{algpseudocode}
\usepackage[linesnumbered,ruled,vlined]{algorithm2e}
\usepackage{nomencl}

\SetCommentSty{mycommfont}
\usepackage{subcaption}

\definecolor{GreenForest}{rgb}{0.09, 0.45, 0.27}
\usepackage{soul}
\usepackage{balance}
\usepackage{cite}
\usepackage{amssymb}
\usepackage{acro} 
\usepackage{steinmetz}
 \DeclareAcronym{6G}{short = 6G, long  = sixth generation }
\DeclareAcronym{5G}{short = 5G, long  = 5th generation }
\DeclareAcronym{ISAC}{short = ISAC , long  = integrated sensing and communication}
\DeclareAcronym{MMSE}{short = MMSE , long  = minimum mean squared error }
\DeclareAcronym{RMSE}{short = RMSE , long  = root mean squared error }
\DeclareAcronym{DFT}{short = DFT, long  = discrete Fourier transform }
\DeclareAcronym{IDFT}{short = IDFT, long  = inverse discrete Fourier transform }
\DeclareAcronym{FFT}{short = FFT, long  = fast Fourier transform }
\DeclareAcronym{ISFFT}{short = ISFFT, long  = inverse symplectic finite Fourier transform }
\DeclareAcronym{IFFT}{short = IFFT, long  = inverse fast Fourier transform }
\DeclareAcronym{SFFT}{short = SFFT, long  = symplectic finite Fourier transform }
\DeclareAcronym{V2X}{short = V2X, long  = vehicle-to-everything }
\DeclareAcronym{NTN}{short = NTN, long  = non-terrestrial networks }
\DeclareAcronym{MSE}{short = MSE, long  = mean square error }
\DeclareAcronym{CSI}{short = CSI , long  = channel state information }
\DeclareAcronym{MIMO}{short = MIMO, long  = multiple-input multiple-output }
\DeclareAcronym{BW}{short = Bw, long  = bandwidth }
\DeclareAcronym{LS}{short = LS, long  = least square }
\DeclareAcronym{SNR}{short = SNR, long  = signal-to-noise ratio }
\DeclareAcronym{NMSE}{short = NMSE, long  = normalized mean square error }
\DeclareAcronym{PAPR}{short = PAPR, long  = peak to average power ratio }
\DeclareAcronym{ISI}{short = ISI, long  = inter-symbol interference }
\DeclareAcronym{ICI}{short = ICI, long  = inter-carrier interference }
\DeclareAcronym{CP}{short = CP, long  = cyclic prefix }
\DeclareAcronym{CPP}{short = CPP, long  = chirp-periodic prefix }
\DeclareAcronym{ZP}{short = ZP, long  = zero padding }
\DeclareAcronym{ZS}{short = ZS, long  = zero suffix }
\DeclareAcronym{ZF}{short = ZF, long  = zero forcing }
\DeclareAcronym{RZP}{short = RZP, long  = reduced-zero padded }
\DeclareAcronym{BER}{short = BER, long  = bit error rate }
\DeclareAcronym{SIC}{short = SIC, long  = successive interference cancellation }
\DeclareAcronym{AWGN}{short = AWGN, long  = additive white Gaussian noise}
\DeclareAcronym{SINR}{short = SINR, long  = signal-to-interference-plus-noise ratio }\DeclareAcronym{SIR}{short = SIR, long  = signal-to-interference ratio ,tag = abbrev}
\DeclareAcronym{BPSK}{short = BPSK, long  = binary phase shift keying }
\DeclareAcronym{QPSK}{short = QPSK, long  = quadrature phase shift keying }
\DeclareAcronym{PDF}{short = PDF, long  = probability density function }
\DeclareAcronym{1D}{short = 1D, long  = 1-dimensional }
\DeclareAcronym{2D}{short = 2D, long  = 2-dimensional }
\DeclareAcronym{3D}{short = 3D, long  = three-dimensional }
\DeclareAcronym{QAM}{short = QAM, long  = quadrature amplitude modulation }
\DeclareAcronym{CE}{short = CE, long  = channel estimation }
\DeclareAcronym{FMCW}{short = FMCW, long  = frequency modulated continuous wave }
\DeclareAcronym{JSAC}{short = JSAC, long  = joint sensing and communication }
\DeclareAcronym{OTFS}{short = OTFS, long  = orthogonal time-frequency space }
\DeclareAcronym{SE}{short = SE, long  =spectral efficiency}
\DeclareAcronym{MP}{short = MP, long  = message passing }
\DeclareAcronym{OMP}{short = OMP, long  = orthogonal matching pursuit }
\DeclareAcronym{ML}{short = ML, long  = maximum likelihood }
\DeclareAcronym{OFDM}{short = OFDM, long  = orthogonal frequency division multiplexing }
\DeclareAcronym{AFDM}{short = AFDM, long  = affine frequency division multiplexing }
\DeclareAcronym{OCDM}{short = OCDM, long  = orthogonal chirp division multiplexing }
\DeclareAcronym{ODDM}{short = ODDM, long  = orthogonal delay-Doppler division multiplexing}
\DeclareAcronym{FD}{short = FD, long  = full duplex}
\DeclareAcronym{PRI}{short = PRI, long  = pulse repetition interval}
\DeclareAcronym{LTV}{short = LTV, long  = linear time-varying  }
\DeclareAcronym{FrFT}{short = FrFT, long  = fractional Fourier transform  }
\DeclareAcronym{IDAFT}{short = IDAFT, long  = inverse discrete affine Fourier transform}
\DeclareAcronym{DAFT}{short = DAFT, long  = discrete affine Fourier transform}

\DeclareAcronym{AFT}{short = AFT, long  = Affine Fourier transform}
\DeclareAcronym{STFT}{short = STFT, long  = short-time Fourier transform}
\DeclareAcronym{DD}{short = DD-coupling, long  = Doppler-delay coupling}
\DeclareAcronym{SBL}{short = SBL, long  = sparse bayesian learning}
\DeclareAcronym{EPA}{short = EPA, long  = embedded pilot-aided}
\DeclareAcronym{PCTD}{short = PCTD, long  = post-coded time domain}
\DeclareAcronym{MRC}{short = MRC, long  = maximum ratio combiner}
\DeclareAcronym{RRC}{short = RRC, long  = root raised cosine}
\DeclareAcronym{EVM}{short = EVM, long  = error vector magnitude}
\DeclareAcronym{EVA}{short = EVA, long  = extended vehicular A}
\DeclareAcronym{SI}{short = SI, long  = self interference}
\DeclareAcronym{CSC}{short = CSC, long  = circularly shifted chirp}
\DeclareAcronym{LFM}{short = LFM, long  = linear frequency modulated}
\DeclareAcronym{LoRa}{short = LoRa, long  = long range}
\DeclareAcronym{Tx}{short = Tx, long  = transmitting}
\DeclareAcronym{Rx}{short = Rx, long  = receiving}
\DeclareAcronym{RCS}{short = RCS, long  = radar cross section}
\DeclareAcronym{CFO}{short = CFO, long  = carrier frequency offset}
\DeclareAcronym{CPI}{short = CPI, long  = coherent processing interval}
\DeclareAcronym{WSS}{short = WSS, long  = wide sense stationary}
\DeclareAcronym{IID}{short = i.i.d., long  = independent and identically distributed}
\DeclareAcronym{NOMA}{short = NOMA, long  = non orthogonal multiple access}
\DeclareAcronym{RDM}{short = RDM, long  = range-Doppler map}
\DeclareAcronym{LoS}{short = LoS, long  = line of sight}
\DeclareAcronym{NLoS}{short = NLoS, long  = non-line of sight}
\DeclareAcronym{IoT}{short = IoT, long  = internet of things}
\DeclareAcronym{TF}{short = TF, long  = time-frequency}
\DeclareAcronym{3GPP}{short = 3GPP, long  = 3rd Generation Partnership project}
\DeclareAcronym{OTSM}{short = OTSM, long  = orthogonal time sequency modulation}
\DeclareAcronym{DMRS}{short = DMRS, long  = Demodulation Reference Signal}
\DeclareAcronym{NR}{short = NR, long  = new radio}
\DeclareAcronym{LTI}{short = LTI, long  = linear time invariant}
\DeclareAcronym{CFR}{short = CFR, long  = channel frequency response}
\DeclareAcronym{TDL}{short = TDL, long  = tapped delay line}
\DeclareAcronym{CDL}{short = CDL, long  = cluster delay line}
\DeclareAcronym{CIR}{short = CIR, long  = channel impulse response}
\DeclareAcronym{MPCs}{short = MPCs, long  = multi-path components}
\DeclareAcronym{PSD}{short = PSD, long  = power spetcrum density}
\DeclareAcronym{PDP}{short = PDP, long  = power delay profile}
\DeclareAcronym{WSSUS}{short = WSSUS, long  = wide sens stationary uncorrelated scatter}
\usepackage[compact]{titlesec}
\titlespacing{\section}{1.0pt}{*1.0}{*0}
\titlespacing{\subsection}{1.1pt}{*1.1}{*0}
\titlespacing{\subsubsection}{0.3pt}{*0}{*0}
\begin{document}

\title{Channel-Aware Waveform Selection Criteria Across Different Waveform Domains
\\}

\author{
\IEEEauthorblockN{Hamza Haif, Abdelali Arous, and H\"{u}seyin Arslan,
\IEEEmembership{Fellow, IEEE}}
\thanks{The authors H. Haif,  A. Arous, and H. Arslan, are with the Department of Electrical and Electronics Engineering, Istanbul Medipol University, Istanbul, 34810, Turkey, (e-mail: hamza.haif@std.medipol.edu.tr; abdelali.arous@std.medipol.edu.tr;  huseyinarslan@medipol.edu.tr). 
}}

\maketitle

\begin{abstract}
Waveform evaluation for sixth generation (6G) networks has largely relied on sparse and quasi-stationary channel models that enabled mathematical tractability, diversity gains, and Doppler robustness. However, such models obscure the propagation complexity of dense urban environments, high-mobility scenarios, and heterogeneous network deployments. This paper sheds light on a generalized and scalable channel model that incorporates cluster birth-death dynamics, Doppler spectral spreading, time-varying delays, and piecewise local stationarity. Based on this model, the effective input-output relationships of the main 6G waveforms are derived, exposing waveform-dependent interference structures that remain hidden under conventional sparse assumptions. Building on these effective channels, a channel-aware waveform prioritization framework is developed based on delay-Doppler resolvability, stationarity conditions, effective signal-to-interference-plus-noise ratio (SINR), and user equipment (UE) cell distribution. Simulation results under the proposed channel model using 3GPP CDL parameters confirm that affine frequency division multiplexing (AFDM) and orthogonal time frequency space (OTFS) retain their spectral efficiency advantage and path combining gains only under sparse, resolvable, stationarity conditions, whereas orthogonal frequency division multiplexing (OFDM) and discrete Fourier transform-spread (DFT-s)-OFDM can be both tuned to achieve superior reliability and more stable performance under the proposed channel model.
\end{abstract}


\section{Introduction}
\IEEEPARstart{T}{he} technological growth in the upcoming \ac{6G} extends beyond enhancing communication throughput, coverage, latency, and capacity to map new radio applications such as \ac{ISAC} , \ac{V2X} \cite{hong2026integrated}, digital twins, \ac{NTN}, and massive \ac{IoT} deployments \cite{medina20253gpp}. These applications can sense, compute, and act independently of the centralized network, thereby; they have different properties, design requirements, and reveal intrinsic channel properties that were not relevant in previous cellular systems. These applications expose the limitations of conventional channel models that were designed primarily around communication-centric assumptions. Classical models decompose the channel into a small set of static or quasi-static parameters, a fixed number of paths, separable delay and Doppler components, and stationary statistics over the observation window. While such simplifications are adequate when the channel is merely an impairment to be equalized, they become insufficient when the channel itself is the object of inference, as in \ac{ISAC}. In reality, the wireless channel is a dynamic, geometrically structured, and non-stationary process that exhibits a richer set of phenomena. Additionally, these non-\ac{WSSUS} channels are characterized by second-order statistics that vary with time and frequency, ~\cite{matz2005non,matz2003doubly}. For instance, as transceivers and scatterers move through the environment, multipath clusters containing dominant rays with similar physical origins continuously appear and disappear, manifesting as clusters or rays emerging and vanishing across local stationarity regions, in what is known as the birth and death phenomenon \cite{meijerink2014physical,gutierrez2017geometry}. Scatterers enter and leave the observable evolving geometry, making the number and identity of propagation paths time-varying \cite{paier2008non}. This non-stationarity is particularly consequential for \ac{ISAC} and digital twin applications, where a newly detected path must be distinguished from a newly appeared physical object, and where tracking algorithms must handle the varying propagation paths over time \cite{11411756}. Beyond cluster-level structure, the mobility-induced Doppler effect will not necessarily be a single discrete shift but rather a Doppler spectrum \cite{7925795}, a continuous spread of frequencies reflecting the varying aspect of mobility and the angles of the impinging rays \cite{wang2023channel}. Combining both Doppler spectrum and the birth-death dynamics results in delay and Doppler coupling, where they are not independent parameters but are geometrically linked through the motion of scatterers. As a target moves, its path length changes continuously, and the rate of that change manifests as the Doppler shift. In wideband with long coherent processing interval (CPI), this coupling produces range migration, where a target traverses multiple delay bins during the integration period, overruling the separability assumption that underpins conventional two-dimensional matched filtering. In \ac{NTN} scenarios, where relative velocities between satellites and ground terminals are extreme, this coupling becomes one of the dominant channel impairments \cite{11432908}. Moreover, for short observation windows and low mobility, a linear approximation capturing only the first-order delay drift suffices. However, for high-speed platforms or for extended CPI, higher-order delay statistics, including acceleration and trajectory curvature, become essential \cite{9473534}. Time-varying delays cause \ac{ISI} in \ac{OFDM} systems by breaking subcarrier orthogonality within a symbol period \cite{Hlawatsch2011}, and smear the delay-Doppler spread function in sensing, degrading range and velocity resolution unless the trajectory is explicitly accounted for in the signal processing \cite{10769778}.
\par Although the aforementioned channel phenomena have been individually investigated in the literature, they have not been collectively consolidated into a unified and tractable channel model \cite{paier2008non,gomez2022doppler,gutierrez2023envelope,he2024wireless,zhang20236g}. These propagation properties collectively define the true complexity of 6G channels that emerges when sensing, mobility, and heterogeneous network architectures are brought together. Characterizing, modeling, and exploiting these properties is therefore a prerequisite for the joint design of waveforms, signal processing algorithms, and channel estimators. Recently, several waveforms have been proposed for 6G, including \ac{OTFS}, \ac{ODDM}, \ac{AFDM}, \ac{OCDM}, and frequency-modulated (FM)-\ac{OFDM} \cite{9392379, 10562334, b5, 9962325, 9346006}. These waveforms offer high diversity order, robustness against time-varying channels, sparse channel representation, high sensing resolution, and desirable security properties. However, these performance advantages are contingent upon the assumptions of channel sparsity, delay-Doppler separability, limited propagation paths, and resolvability of uncorrelated rays, which renders their evaluation under such assumptions practically unverified. This is further underscored by the \ac{3GPP} standardization decision to retain \ac{OFDM} as the primary waveform for 6G, which raises the question of why these waveforms, despite their superior theoretical performance compared to \ac{OFDM}, have not gained standardization support. A principal reason is that these waveforms have not been evaluated under realistic channel models. While advanced channel models accurately represent the propagation environment, they suffer from high computational complexity, non-stationarity and time-variation modeling challenges, and lack of reproducibility and mathematical tractability. Collectively, these limitations create a bottleneck for waveform evaluation under channel models that are simultaneously realistic, scalable, and mathematically tractable.
\par Based on the above, this paper proposes a generalized, scalable, and mathematically tractable channel model for waveform evaluation, formulated in both continuous and matrix form. Each channel component is investigated and related to its intrinsic propagation properties with mathematical reasoning, while ensuring that the proposed model reduces to the conventional sparse channel model under appropriate simplifying assumptions. Using the proposed channel model, the input-output relationships of the prominent 6G waveforms; \ac{OFDM}, \ac{AFDM}, \ac{OTFS}, and single-carrier signals are derived, highlighting their respective domain-specific effective channel properties. Building on these derivations, an adaptive channel-aware waveform selection framework is proposed, establishing a principled basis for waveform design and selection according to the observed channel conditions and the targeted application. Furthermore, it is demonstrated how the resolution of channel parameters can be leveraged for cell-region-based regime-adaptive waveform selection and its relation to the aggregated received \ac{SNR}. The main contributions of this work are summarized as follows:
\begin{itemize}
    \item A generalized, scalable, and parametric channel model is proposed to reflect real propagation effects, incorporating intrinsic channel properties including the birth and death of multipath clusters, inter-cluster interference, Doppler spectral spread, and time-varying delays. The generation, distribution, and integration of each parameter within the compact model are detailed. Furthermore, stationarity regions for \ac{WSSUS} and non-\ac{WSSUS} propagation are formally defined and linked to the channel parameters.   
    \item The input-output relationships of prominent 6G waveforms including \ac{AFDM}, \ac{OFDM}, \ac{OTFS}, and DFT-s-\ac{OFDM} are derived under the proposed channel model, revealing the distinctive effective channel matrix properties induced by different channel components for each waveform. It is further demonstrated that the derived input-output relationships reduce to their conventional counterparts under appropriate sparsity and resolvability assumptions.
    \item Capitalizing on the derived effective channel matrices, an adaptive channel-dependent waveform selection framework is proposed, jointly accounting for desired channel properties, \ac{SINR}, interference, and achievable rate, providing an optimal waveform design and selection framework supported by comprehensive analysis.   
    \item Finally, a comprehensive performance evaluation of the proposed channel-aware waveform selection scheme is conducted, assessing \ac{BER}, \ac{SE}, \ac{PAPR}, achievable rate, and \ac{CE} \ac{MSE} under varying channel conditions of the proposed channel model. 
\end{itemize}
\footnote{\textit{Notation}: Bold uppercase $\mathbf{A}$, bold lowercase $\mathbf{a}$, and unbold letters $a$ denote matrices, column vectors, and scalar values, respectively. $\mathbf{A}\circ\mathbf{B}$ denotes element-wise multiplication. $[\alpha]_{\beta}$ and $\lfloor a \rfloor$ denote modulo-$\beta$ reduction of $\alpha$ and the floor operation, respectively. $\delta(\cdot)$ denotes the continuous-time Dirac delta function, while $\delta[\cdot]$ denotes the discrete Kronecker delta. $\mathbb{E}\{\cdot\}$ denotes the expectation operator. $\mathbb{Z}^{+}$ and $\mathbb{C}^{M\times N}$ denote the set of positive integers and the space of $M\times N$ complex-valued matrices, respectively. $\mathcal{N}(\mu,\sigma^2)$ and $\mathcal{U}(a,b)$ denote the real Gaussian distribution with mean $\mu$ and variance $\sigma^2$, and the uniform distribution, respectively. The symbol $j^2=-1$ denotes the imaginary unit.}
\begin{figure*}[t]
    \centering
    \includegraphics[width=\textwidth, height=0.28\textheight]{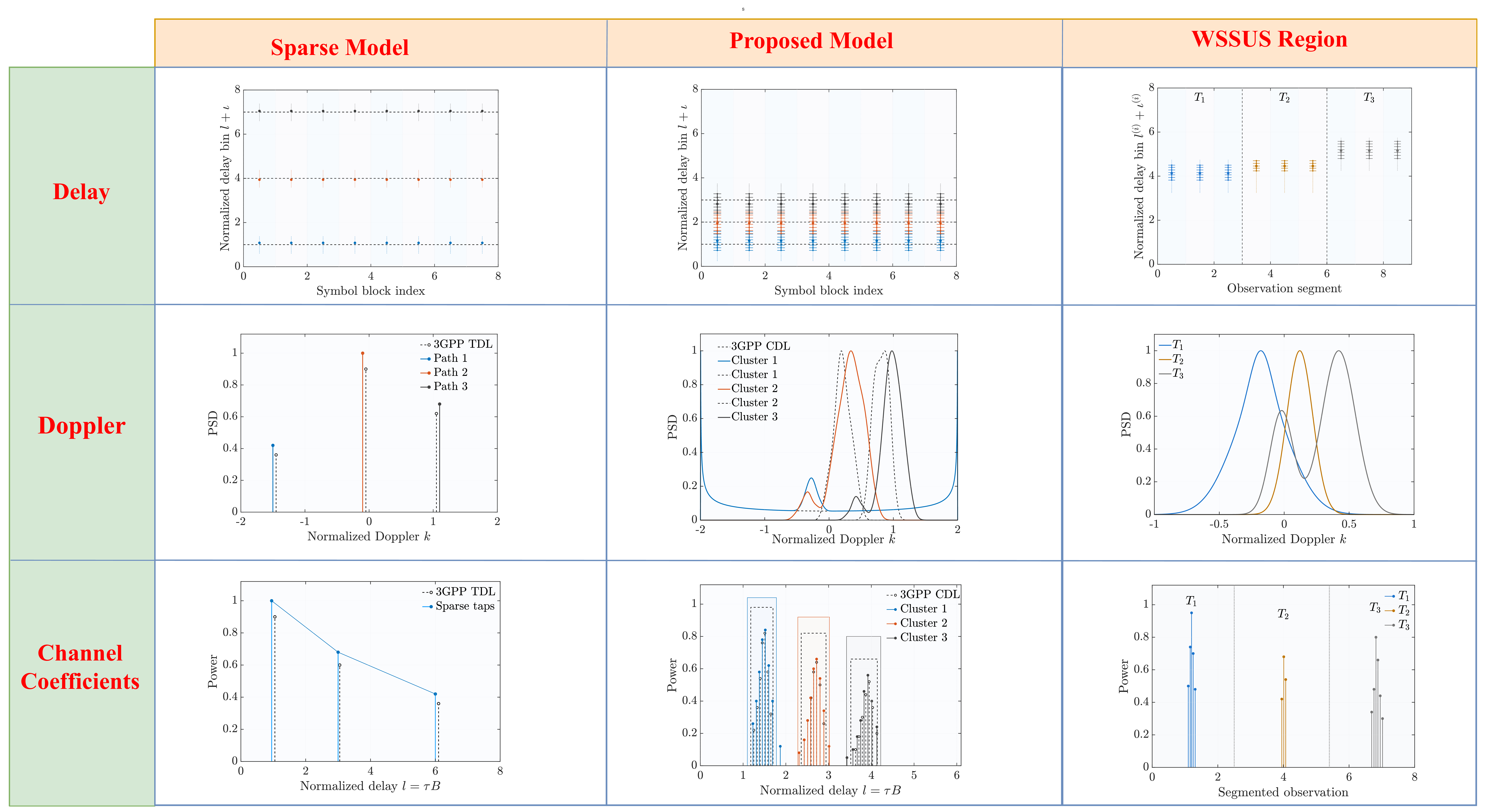}
    \caption{Behavior of channel effects: delays, Doppler shifts, and channel coefficients under sparse channel model and proposed channel model and extended to a single tap evolution within and beyond WSSUS region.}
    \label{fig:block_diag}
\end{figure*}
\section{Scalable Channel Model}
This section dives into the analysis of a parametric channel model that is used to capture the stationarity behavior under realistic channel propagation assumptions. Specifically, the captured mobility scenarios discussed in \ac{6G} standardization \cite{yang2026channel}. The model serves as a baseline framework upon which the 6G waveforms will be analyzed. \\
The \ac{CIR} is modeled using a $C$ dominant clusters characterized by a  varying number of rays $R_c$ within each cluster, with each ray linked to an attenuation factor $h_{c,r}$. Every cluster shares a common delay $\tau_c$. For the Doppler $\nu_{c,r}$, it follows a spectrum shape coming from a single shift corresponding to each ray within the cluster. The model is given as
\begin{equation}
    h(t,\tau) = \sum^{C}_{c=1} \sum^{R_c(t)}_{r=1} h_{c,r}e^{j2\pi \nu_{c,r}(t-\tau_{c}(t))}\delta(\tau-\tau_{c}(t)).
    \label{eq:Channel_model}
\end{equation}
The model presented in \eqref{eq:Channel_model} aims to provide an approximation to the real environment propagation independent of the transmitted signal, while further discussion on this is going to be presented in Section. II.B.
\subsection{Channel Parameters}
In this subsection, the structure of \ac{MPCs}, the delay process behavior, and the Doppler \ac{PSD} of the model in \eqref{eq:Channel_model} are analyzed, while establishing the general formula for generating each parameter.
\subsubsection{Multi-Path Components}
In a rich scattering environment, the arrival of \ac{MPCs} tends to be correlated and impinging from different angles, known as \ac{CDL}. Each cluster is a combination of a group of rays that represent the reflecting points of a superscatter. Let $C$ denote the number of delay clusters within a channel realization where based on \cite{meijerink2014physical}, it follows a Poisson distribution $C \sim \mathrm{Poisson}(\lambda_C)$, with $\lambda_C$ being the Poisson arrival rate of cluster delays, where $\lambda_C$ is governed by the richness of the environment and the propagation scenario, with \ac{NLoS} tending to have larger $\lambda_C$ compared to \ac{LoS} \cite{paier2008non}. Depending on $C$, the rays also follow a Poisson distribution, such as $R_c \sim \mathrm{Poisson}(\lambda_R)$ where $\lambda_R$ is the Poisson arrival rate of rays within each cluster, which describes the density of the intra-cluster MPCs.\\
Next, to generate $h_{c,r}$, a more realistic approach is followed instead of the conventional exponentially decaying \ac{PDP}. Let $X_c\sim\mathcal{N}(0,\sigma_{\mathrm{sh}}^2)$ be the
log-normal shadowing term, and $g(\tau_c)$ is the envelope \ac{PDP}, with $\sigma_{\mathrm{sh}}^2$ as the shadowing variance. In the exponential case,
$g(\tau_c)=\exp(\frac{-\tau_c}{\tau_{\mathrm{RMS}}})$, where $\tau_{\mathrm{RMS}}$ is the root mean square delay spread of the \ac{PDP}. The cluster power is
then \cite{molisch2012wireless}
\begin{equation}
P_c = \frac{\tilde{P}_c}{\sum_{c'=1}^{C}\tilde{P}_{c'}},
\qquad
\tilde{P}_c \triangleq g(\tau_c)\,10^{X_c/10},
\label{eq:cluster_power}
\end{equation}
where the normalization ensures $\sum_c P_c = 1$.\\
After that, the cluster power $P_c$ is divided among $R_c$ rays to generate the individual ray's channel coefficient such as
\begin{equation}
h_{c,r}=|h_{c,r}|e^{j\phi_{c,r}}=\sqrt{P_c\,w_{c,r}}\,e^{j\phi_{c,r}},
\label{eq:hcr_def}
\end{equation}
where $w_{c,r}$ is the normalized ray-power fraction defined as
\[
w_{c,r}=\frac{u_{c,r}}{\sum_{m=1}^{R_c}u_{c,m}},
\qquad
u_{c,r}\sim\mathrm{Exp}(1),
\]
and $\phi_{c,r}$ is the random phase, with $\phi_{c,r}\sim\mathcal{U}[0,2\pi)$.
\subsubsection{Propagation Delays}
Channel delays are described on the basis of the geometric evolution of the reflector with respect to the transmitter and receiver. Let $d_{c,r}(t)$ denote the time-varying propagation distance of ray $(c,r)$, and let
\begin{equation}
\tau_{c,r}(t)=\frac{d_{c,r}(t)}{c_0},
\end{equation}
where $c_0$ is the speed of light. At the geometric level, ray delays may differ within the same cluster. However, at the resolvable signal level, all rays within cluster $c$ are represented by the common cluster delay $\tau_c(t)$ used in \eqref{eq:Channel_model}. Under relative radial motion, the ray delay at reference $t_0$ evolves as \cite{matz2005non}
\begin{equation}
\begin{aligned}
d_{c,r}(t)&=d_{c,r}(t_0)+v_{c,r}(t_0)(t-t_0),
\\
\tau_{c,r}(t)&=\tau_{c,r}(t_0)+\beta_{c,r}(t-t_0),
\end{aligned}
\end{equation}
with $\beta_{c,r}\triangleq v_{c,r}(t_0)/c_0$ denoting the delay drift induced by mobility. Consequently, delay and Doppler parameters must be jointly generated to preserve their geometric coupling, an independence assumption being valid only within locally stationary regions, as discussed in Section~II-B.
\subsubsection{Doppler Shifts}
Doppler shift $f_{D}= \frac{v}{c_0}f_c$ originates from the variation of the propagation distance due to the mobility of the transmitter, receiver or the reflector mobility, where $v$ is the receiver speed magnitude and $f_c$ is the carrier frequency. For ray $(c,r)$ arriving with azimuth $\theta_{c,r}$ relative to the velocity direction $\theta_v$, the Doppler is
\begin{equation}
\nu_{c,r}= f_{D}\cos(\theta_{c,r}-\theta_v).
\end{equation}
To represent realistic scattering, the Von Mises-Fisher model is adopted as an angle driven Doppler spectrum model \cite{gutierrez2023envelope}, where the impinging angles are random with a scenario dependent power azimuth spectrum. Specifically, for each cluster $c$, the ray angles are drawn according to a distribution $p_c(\theta)$ supported on a sector $\left[\bar{\theta}_c-\Delta\theta_c/2,\ \bar{\theta_c}+\Delta\theta_c/2\right]$, where $\bar{\theta_c}$ is the mean angle of cluster or a mixture of sectors for multi-lobe scattering and the resulting Doppler spectrum is also determined with the ray power weights $w_{c,r}$. When the scattering is isotropic, $p_c(\theta)=\mathcal{U}[0,2\pi)$, the induced Doppler spectrum reduces to the classical Clarke-Jakes spectrum.
\subsection{Stationarity Regions}
The channel model in \eqref{eq:Channel_model} is defined to capture the stationary evolution of the channel under mobility. Therefore, stationarity region based interpretation for the WSSUS and non-WSSUS is adopted.
\subsubsection{WSSUS and non-WSSUS conditions}
Let $L_H(t,f)$ denote the time-frequency transfer function
\begin{equation}
L_H(t,f)=\int h(t,\tau)e^{-j2\pi f\tau}\,d\tau .
\label{eq:LH_def}
\end{equation}
In the conventional \ac{WSSUS} model, the channel is wide-sense stationary in $t$ and exhibits uncorrelated scattering in $\tau$:
\begin{equation}
\mathbb{E}\!\left\{h(t,\tau)\,h^{\ast}(t-\Delta t,\tau-\Delta\tau)\right\}
=
r_H(\Delta t,\tau)\,\delta(\Delta\tau).
\label{eq:wssus_def}
\end{equation}
Practical mobile channels, however, are generally non-\ac{WSSUS}, so their second-order statistics vary over time and frequency. Following \cite{matz2003doubly}, this behavior is described through the local scattering function $C_H(t,f;\tau,\nu)$, for which
\begin{equation}
C_H(t,f;\tau,\nu)\neq C_H(t',f';\tau,\nu),
\qquad
(t,f)\neq(t',f').
\label{eq:nonwssus_lsf}
\end{equation}
Although the channel is globally nonstationary, it may remain approximately stationary within short intervals $\mathcal{T}_i=[t_i,t_{i+1})$. Thus, the local scattering function within $\mathcal{T}_i$ satisfies
\begin{equation}
\max_{t\in\mathcal{T}_i}
\left|
C_H(t_i,f;\tau,\nu)-C_H(t_{i+1}^{-},f;\tau,\nu)
\right|
\le \varepsilon,
\label{eq:local_stationarity_lsf}
\end{equation}
where $\varepsilon$ is a small tolerance factor. In the channel model of \eqref{eq:Channel_model}, violations of \eqref{eq:local_stationarity_lsf} arise from the evolution of $\tau_c(t)$, $\nu_{c,r}$, and $|h_{c,r}(t)|^2$ across successive regions. Within $\mathcal{T}_i=[t_i,t_{i+1}), \text{ } i \in \mathcal{I}$ $\mathcal{I}$, where is the number of segments, the ray count remains constant, i.e.,
\begin{equation}
R_c(t)=R_{c,i}, \quad t\in\mathcal{T}_i,
\qquad 
R_{c,i+1}\neq R_{c,i}\ \text{at}\ t=t_{i+1}.
\label{eq:Rc_birth_death}
\end{equation}
Let $\mathcal{R}_{c,i}$ denote the set of active rays of cluster $c$ within region $\mathcal{T}_i$. Then the effective cluster coefficient is
\begin{equation}
h_c^{(i)}(t) \triangleq \sum_{r\in\mathcal{R}_{c,i}} h_{c,r}^{(i)} e^{j2\pi \nu_{c,r}^{(i)}(t-\tau_c^{(i)})},
\qquad t\in\mathcal{T}_i.
\label{eq:cluster_coeff_region}
\end{equation}
Varying $R_{c,i}$ changes both the channel gain and the Doppler spectrum associated with delay $\tau_c^{(i)}$, showing that time selectivity may arise from birth death dynamics in addition to Doppler. Fig. \ref{fig:block_diag} Illustrates the behavior differences of the channel coefficient, delay, and Doppler shift between the sparse channel model and the proposed model, while also showcasing the evolution of these parameters within and beyond the WSSUS region.
\begin{figure*}
\begin{equation} \small \tag{46}
\begin{aligned}
H^{\mathrm{dd}}_{\mathrm{eff}}\!\big[(\mu,a),(\nu,b)\big]
=\sum_{i}^{\mathcal{I}}\sum_{c=1}^{C_i}\sum_{r=1}^{R_{c,i}}
\frac{h_{c,r}^{(i)}}{N'N}
\sum_{l=0}^{N'-1}\sum_{l'=0}^{N'-1} 
\mathbf{1}_{\{a+l M'\in\mathcal{T}_i\}}
 e^{-j2\pi\frac{\mu l}{N'}}\;e^{+j2\pi\frac{\nu l'}{N'}}
\exp\!\left(j2\pi\frac{k_{c,r}^{\mathrm{tot}}}{N}\big(a+l M'\big)\right) \\
\times \sum_{u=0}^{N-1}\exp\!\left(j2\pi\frac{u}{N}\big((a-b)+M'(l-l')-l_{c}^{\mathrm{tot},i}\big)\right).
\end{aligned}
\label{eq:Heff_otfs_final}
\end{equation}
\begin{equation} \small \tag{51}
\begin{aligned}
H^{\mathrm{DFT}}_{\mathrm{eff}}[m,n]
=\sum_{i=1}^{\mathcal{I}}\sum_{c=1}^{C_{i}}\sum_{r=1}^{R_{c,i}}
\frac{h_{c,r}^{(i)}}{N^{2}N_{d}}&
\sum_{u=0}^{N-1}\exp\!\left(-j2\pi\frac{u}{N}l_{c}^{\mathrm{tot},i}\right) 
\sum_{p\in\mathcal{N}_{i}}\sum_{q=0}^{N-1}
\sum_{a=0}^{N_{d}-1}\exp\!\left(-j2\pi\frac{m}{N_{d}}a\right)\exp\!\left(-j2\pi\frac{k_{0}+a}{N}p\right)
\\ & \times \sum_{b=0}^{N_{d}-1}\exp\!\left(+j2\pi\frac{n}{N_{d}}b\right) \times \exp\!\left(+j2\pi\frac{k_{0}+b}{N}q\right) 
\exp\!\left(j2\pi\frac{k_{c,r}^{\mathrm{tot}}+u}{N}p\right)
\exp\!\left(-j2\pi\frac{u}{N}q\right).
\end{aligned}
\label{eq:Heff_sc_final}
\end{equation}
\hrulefill
\end{figure*}
\subsection{Proposed Channel Model in Matrix Form}
Consider the sampled channel over an observation window of duration $T$, represented on a discrete grid of $N$ time samples. The window spans a signal bandwidth of $B$, where $B$ and $T$ are treated as independent design parameters dictated by the s ignal waveform such that $T \neq 1/B$.  \\
To preserve a scalable yet tractable description under mobility, an interpretation of the stationarity region is adopted such that $C$ and $R_c$  within the stationarity interval $\mathcal{T}_i=[n_i,n_{i+1})$ are defined as
\begin{equation}
C=C_i,\qquad R_c(n)=R_{c,i}, \qquad \forall n\in\mathcal{T}_i,
\end{equation}
where $C_i$ is the number of resolvable delay clusters and $R_{c,i}$ is the number of rays within cluster $c$. Across region boundaries, $(C_i,\{R_{c,i}\})$ may change due to the birth and death phenomena described in Section~II.B.\\
Contemplating finite bandwidth resolution, the normalized delays associated with cluster $c$ and its rays are defined as
\begin{equation}
l_c^{\mathrm{tot}} \triangleq \tau_c B = l_c + \iota_c,
\label{eq:delay_norm_B}
\end{equation}
where $l_c^{\mathrm{tot}}$ is the total normalized delay containing both integer and fractional parts, $l_c\in\mathbb{Z}$ is the integer delay part, and $\iota_c\in\left(-\tfrac{1}{2},\tfrac{1}{2}\right]$ is the fractional delay. At the ray level, $\tau_{c,r}=\tau_c+\delta\tau_{c,r}$ and, due to the delay resolution $1/B$, rays within the same cluster satisfy $|\delta\tau_{c,r}|\le \tfrac{1}{2B}$, hence
\begin{equation}
l_{c,r}^{\mathrm{tot}} \triangleq \tau_{c,r}B = l_c^{\mathrm{tot}} + \delta l_{c,r},
 \text{ where  } \delta l_{c,r}\triangleq \delta\tau_{c,r} B,
\label{eq:delay_norm_B_ray}
\end{equation}
which confirms that all rays of cluster $c$ fall within the same resolvable delay bin, unless an ultra wide bandwidth is used like FR2 and subTHz bands.\\
Similarly, the normalized Doppler shift of ray $(c,r)$ is defined with respect to the observation duration $T$ as
\begin{equation}
k_{c,r}^{\mathrm{tot}} \triangleq T\nu_{c,r} = k_{c,r} + \kappa_{c,r},
\label{eq:doppler_norm_T}
\end{equation}
where $k_{c,r}\in\mathbb{Z}$ and $\kappa_{c,r}\in\left(-\tfrac{1}{2},\tfrac{1}{2}\right]$ denote the integer and fractional Doppler indices, respectively.\\
Let $\mathbf{F}\in\mathbb{C}^{N\times N}$ denote the unitary DFT matrix with entries
\begin{equation}
[\mathbf{F}]_{m,n}=\frac{1}{\sqrt{N}}e^{-j2\pi \frac{mn}{N}},\qquad m,n=0,\ldots,N-1.
\end{equation}
The circular fractional delay shift operator associated with a (possibly fractional) normalized delay $l$ is defined via a frequency domain linear phase ramp as
\begin{equation}
\mathbf{P}(l)\triangleq \mathbf{F}^{H}\,\mathbf{\Lambda}(l)\,\mathbf{F},
\qquad 
\mathbf{\Lambda}(l)\triangleq \mathrm{diag}\!\left(e^{-j2\pi \frac{m}{N}l}\right)_{m=0}^{N-1}.
\label{eq:frac_delay_operator}
\end{equation}
The resolvable delay is treated as constant within each stationarity region and may vary only across regions. Therefore, the delay is indexed by the region-wise index $l_c^{\mathrm{tot},i}$. Similarly, the Doppler modulation associated with a (possibly fractional) normalized Doppler $k$ is represented by the diagonal operator
\begin{equation}
\mathbf{D}(k)\triangleq \mathrm{diag}\!\left(e^{j2\pi \frac{k}{N}n}\right)_{n=0}^{N-1}.
\label{eq:doppler_operator}
\end{equation}
Within $\mathcal{N}_i$ where $\mathcal{N}_i=[n_i,n_{i+1})$, the channel over the full frame length is represented by
\begin{equation}
\mathbf{H}_i = \sum_{c=1}^{C_i}\sum_{r=1}^{R_{c,i}} h_{c,r}^{(i)}\,\mathbf{D}\!\left(k_{c,r}^{\mathrm{tot}}\right)\,\mathbf{P}\!\left(l_c^{\mathrm{tot},i}\right).
\label{eq:channel_matrix_region_timevarying}
\end{equation}
The overall channel operator over the full frame is denoted by $\mathbf{H}\in\mathbb{C}^{N\times N}$ and is obtained by selecting, for each time index, the region-wise operator corresponding to the active stationarity region. Using the diagonal selector matrix
\begin{equation}
\mathbf{W}_i \triangleq \mathrm{diag }(\mathbf{1}_{\{n\in\mathcal{T}_i\}}\big)_{n=0}^{N-1}.
\label{eq:Wi_def}
\end{equation}
the time domain channel matrix is expressed compactly by contaminating  $\mathbf{H}_i$ across $\mathcal{I}$ stationarity regions, which results in
\begin{equation}
\mathbf{H} = \sum_{i=1}^{\mathcal{I}} \mathbf{W}_i\,\mathbf{H}_i.
\label{eq:H_global_sum}
\end{equation}
The delay operator $\mathbf{P}(l)$ in \eqref{eq:frac_delay_operator} is circulant under the assumption of a \ac{CP} of length $N_{\mathrm{cp}}$, provided $N_{\mathrm{cp}} \gg l_{\max}$. Consequently, each region-wise matrix $\mathbf{H}_i$ is circulant. The global matrix $\mathbf{H}$ in \eqref{eq:H_global_sum}, however, is only piecewise circulant due to the region-selection matrices $\mathbf{W}_i$. 
\section{Generalized Input-Output relationship}
In this section, the input-output relations and effective channel matrices of the considered waveforms are derived while highlighting the respective domain-specific channel properties.
\begin{figure*}
    \centering
    \begin{subfigure}[b]{0.2\textwidth}
        \centering
        \includegraphics[width=\textwidth]{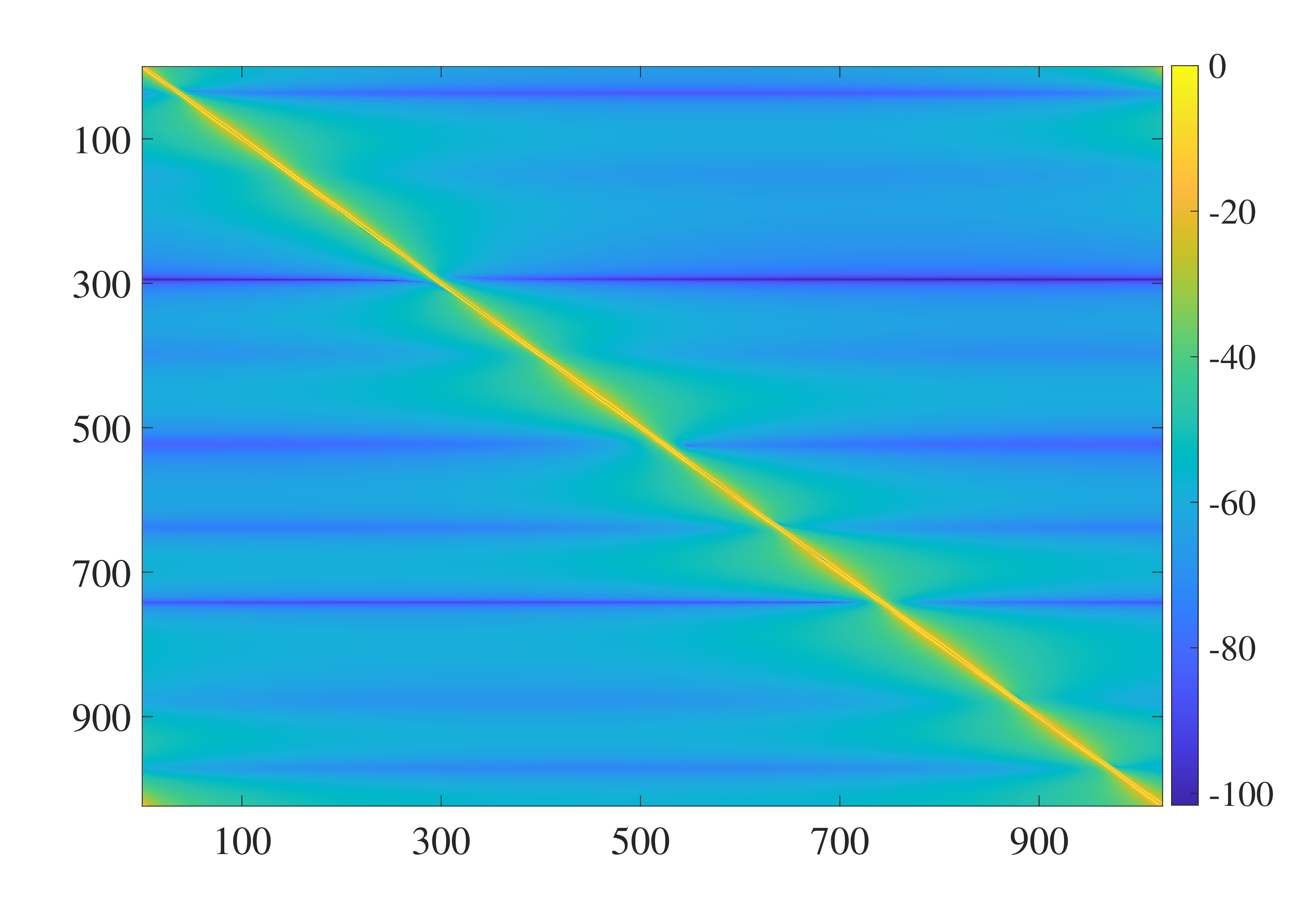}
        \caption{OFDM}
        \label{fig:affine_3tap}
    \end{subfigure}\hfill
    \begin{subfigure}[b]{0.2\textwidth}
        \centering
        \includegraphics[width=\textwidth]{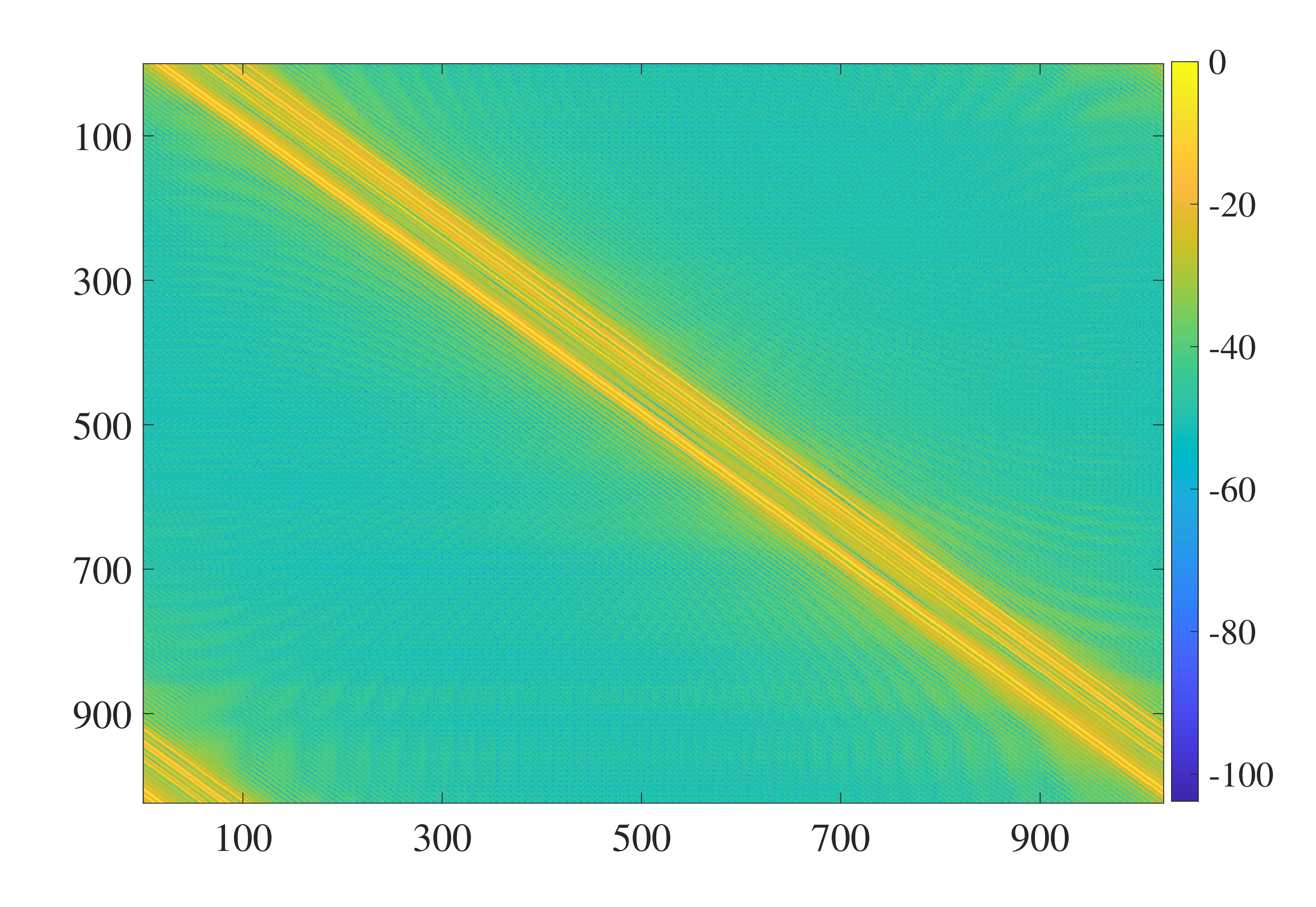}
        \caption{AFDM}
        \label{fig:affine_Ntap}
    \end{subfigure}\hfill
    \begin{subfigure}[b]{0.2\textwidth}
        \centering
        \includegraphics[width=\textwidth]{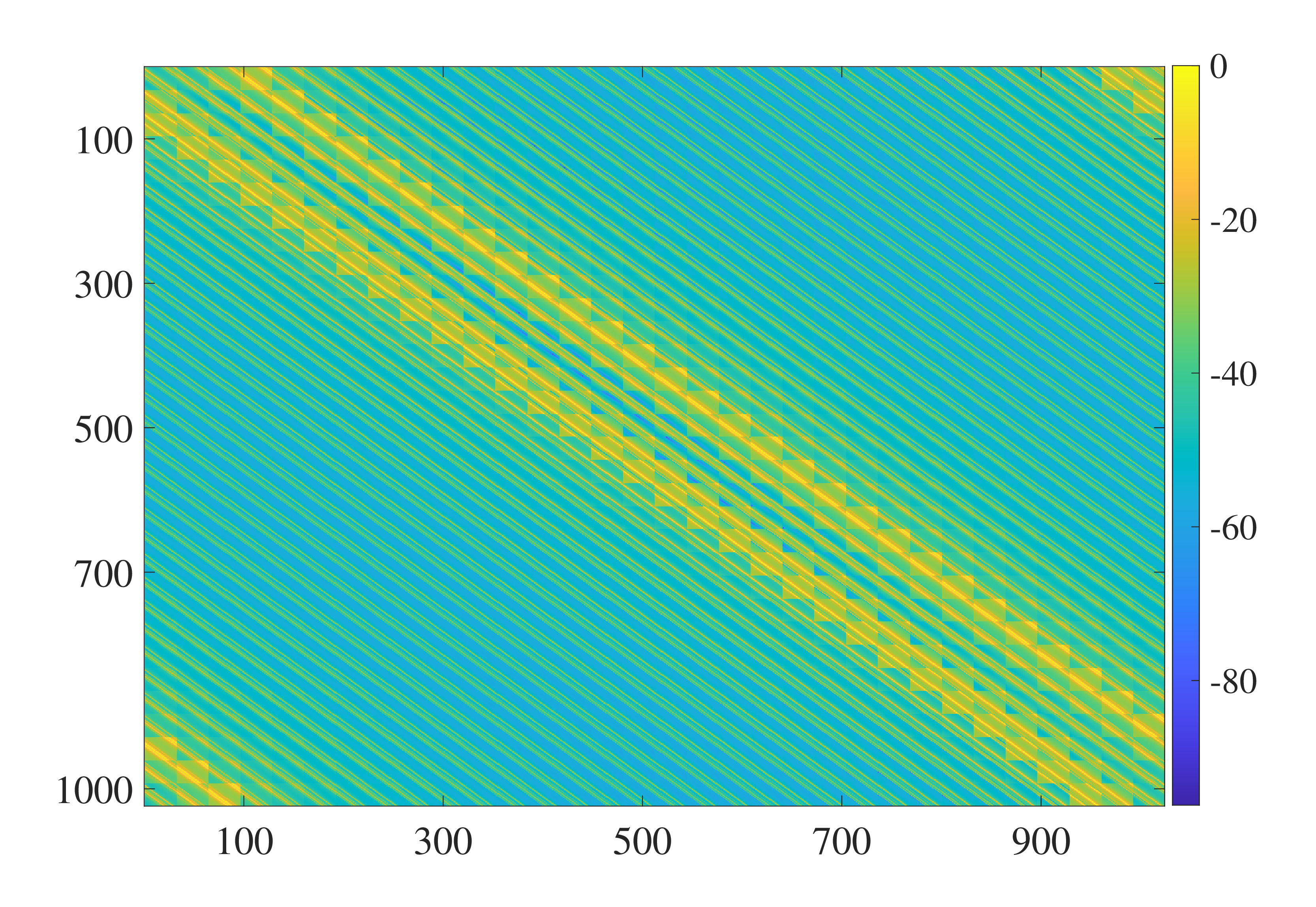}
        \caption{OTFS}
        \label{fig:affine_DD}
    \end{subfigure}\hfill
  \begin{subfigure}[b]{0.2\textwidth}
        \centering
        \includegraphics[width=\textwidth]{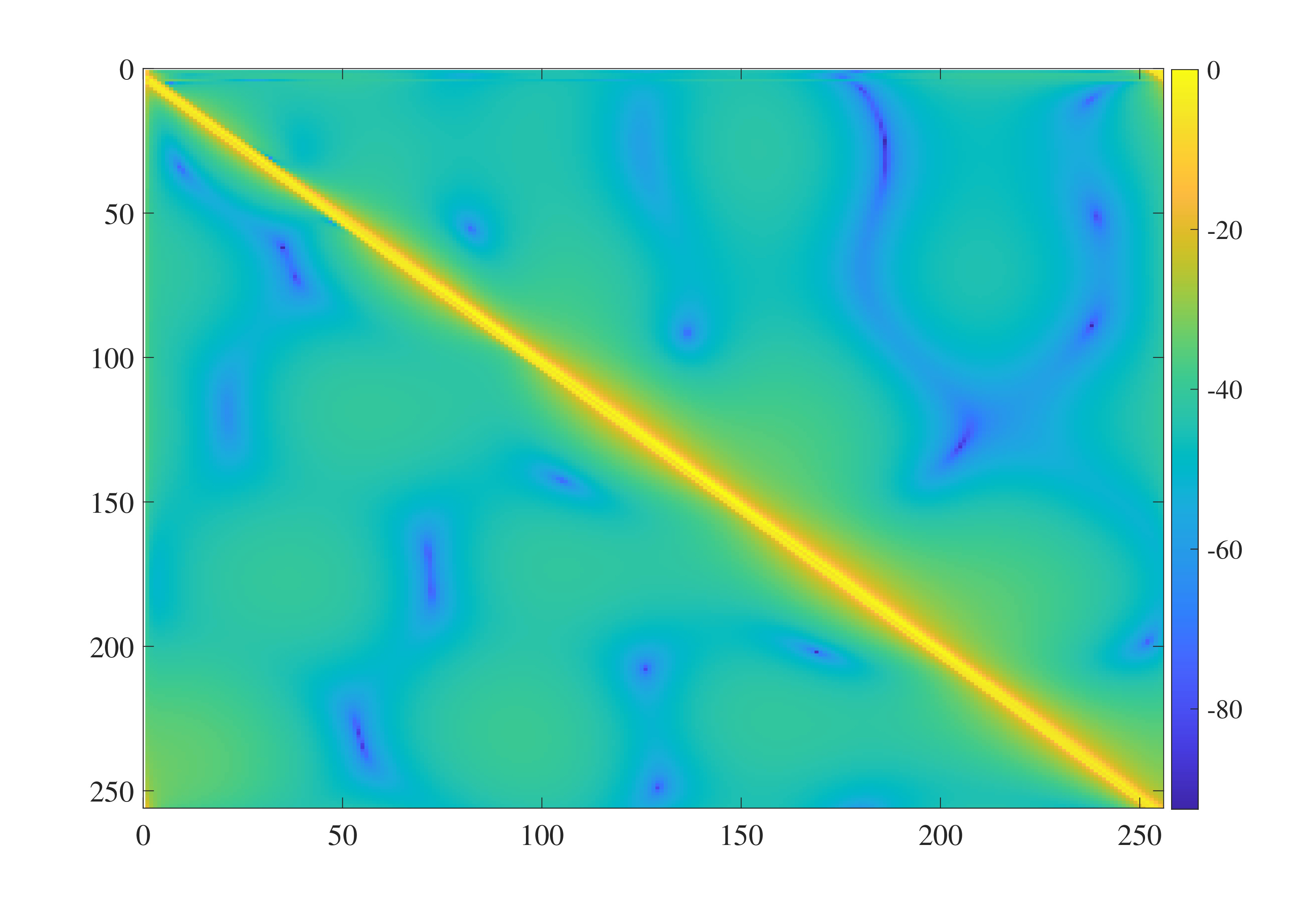}
        \caption{DFT-s-OFDM}
        \label{fig:dft_spa}
    \end{subfigure}
    \vspace{0.8em}

    \begin{subfigure}[b]{0.2\textwidth}
        \centering
        \includegraphics[width=\textwidth]{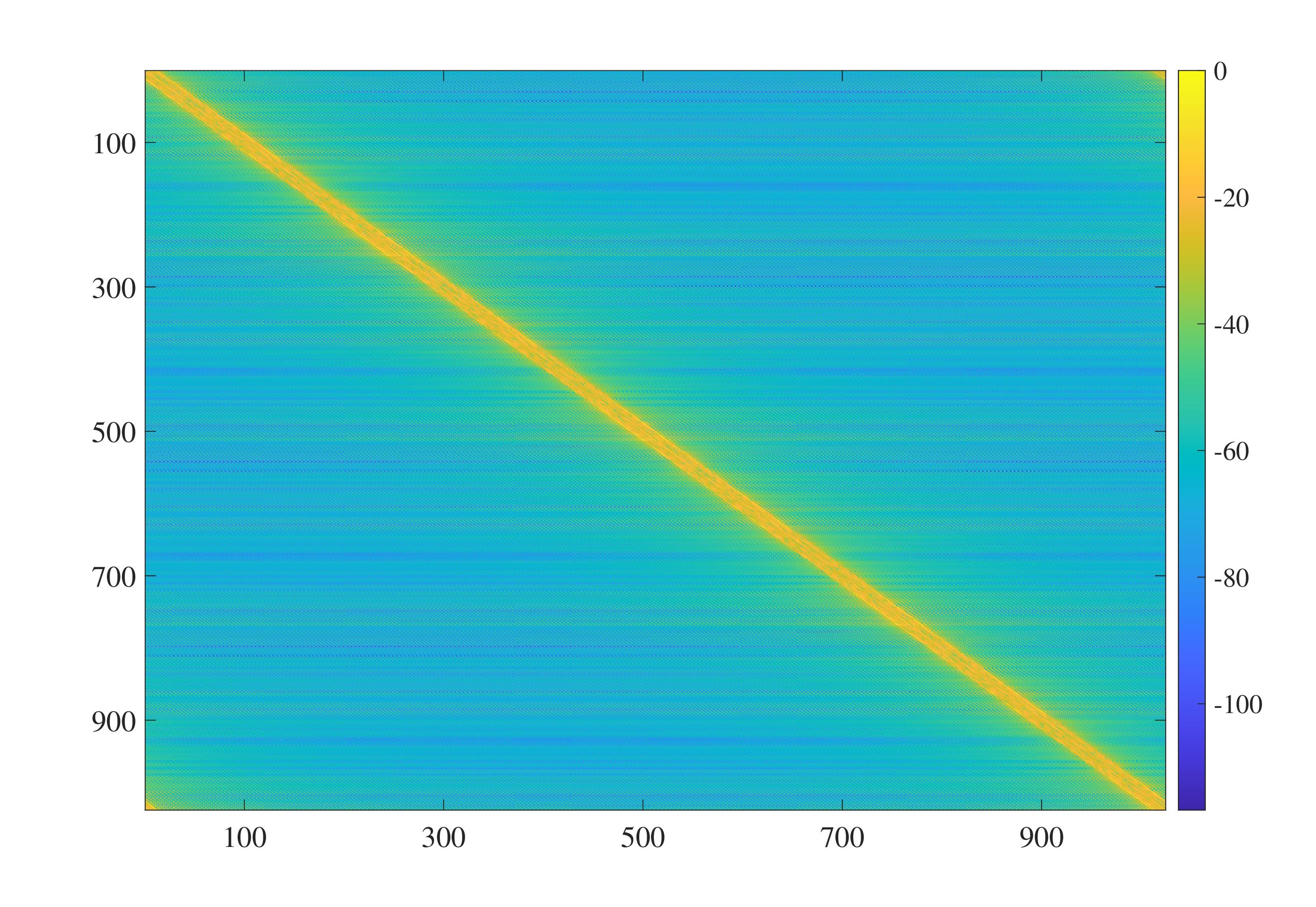}
        \caption{OFDM}
        \label{fig:freq_3tap}
    \end{subfigure}\hfill
    \begin{subfigure}[b]{0.2\textwidth}
        \centering
        \includegraphics[width=\textwidth]{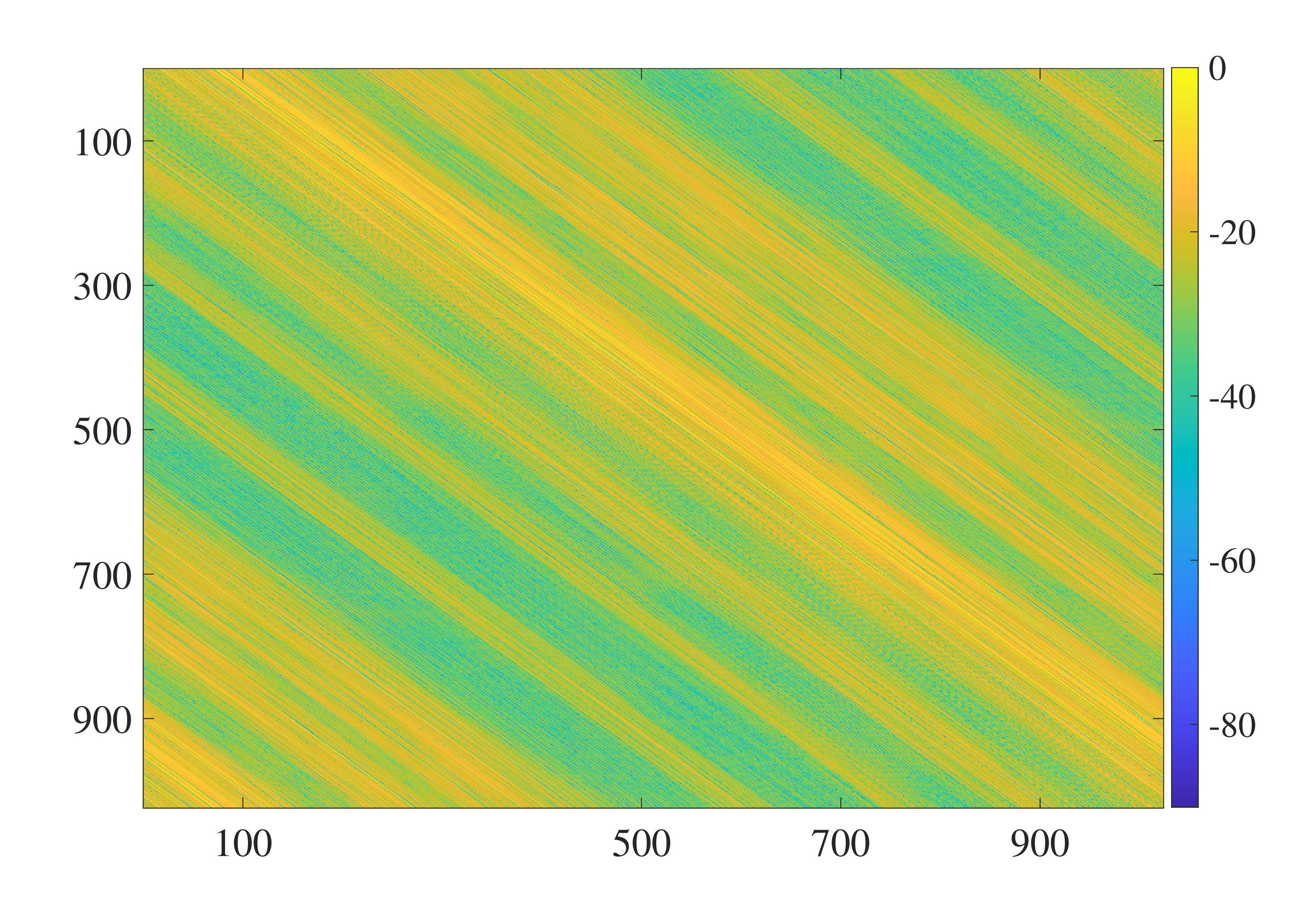}
        \caption{AFDM}
        \label{fig:freq_Ntap}
    \end{subfigure}\hfill
    \begin{subfigure}[b]{0.2\textwidth}
        \centering
        \includegraphics[width=\textwidth]{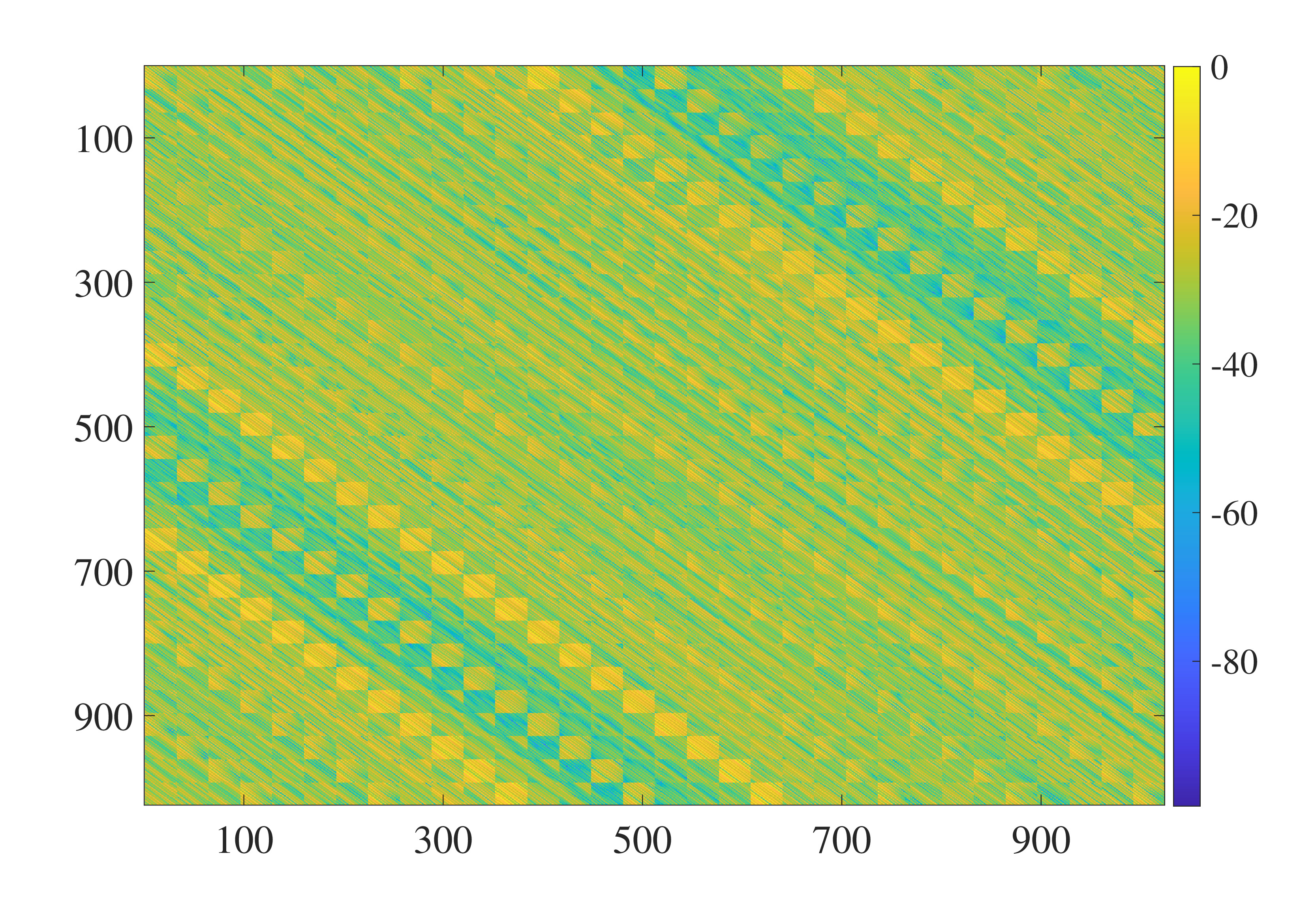}
        \caption{OTFS}
        \label{fig:freq_DD}
    \end{subfigure}\hfill
    \begin{subfigure}[b]{0.2\textwidth}
        \centering
        \includegraphics[width=\textwidth]{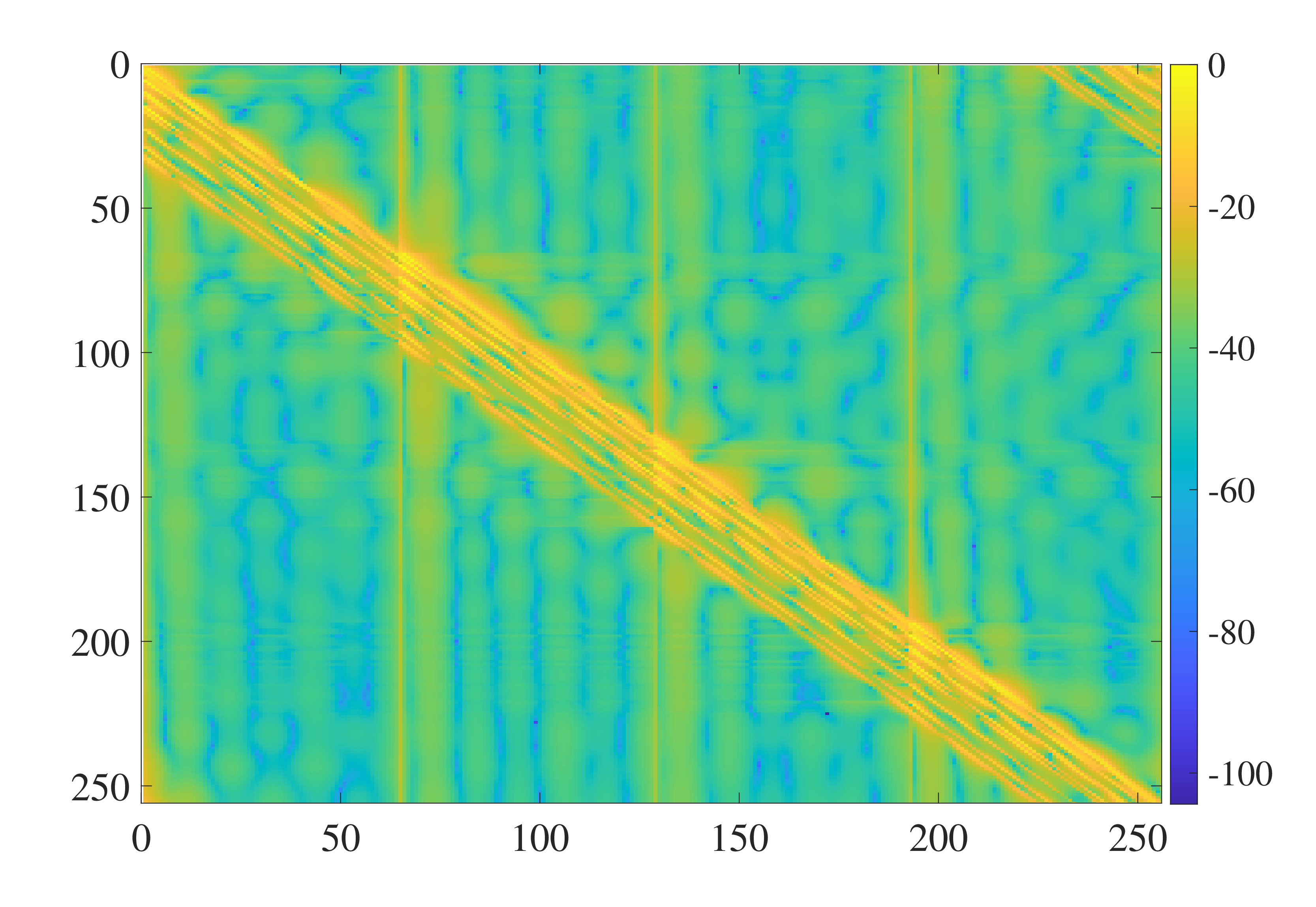}
        \caption{DFT-s-OFDM.}
        \label{fig:dft_adv}
    \end{subfigure}
    \caption{Effective channel representation across OFDM, DFT-s-OFDM, AFDM, and OTFS waveforms: The first row represents the sparse model with $\Delta f = 15$KHz for OFDM and DFT-s-OFDM and $\Delta f = 1$KHz for AFDM and OTFS. The second row represents the sparse model with $\Delta f = 60$ kHz for OFDM and DFT-s-OFDM and $\Delta f = 0.5$ kHz for AFDM and OTFS. }
    \label{fig: CH_eff_representation} 
\end{figure*}
\subsection{AFDM Waveform}
The input-output relation between the AFDM transmitted vector $\mathbf{x}_{\text{af}}$ and the received vector $\mathbf{y}_{\text{af}}$ is given as 
\begin{equation}
\mathbf{y}_{\mathrm{af}}=\boldsymbol{\Phi}\,\mathbf{H}\,\boldsymbol{\Phi}^{H}\mathbf{x}_{\mathrm{af}}+\mathbf{w},
\label{eq:I/O afdm}
\end{equation}
where $\mathbf{w}\sim\mathcal{CN}(\mathbf{0},\sigma^2\mathbf{I})$ is \ac{AWGN}, whose statistics are preserved since the \ac{DAFT} matrix $\boldsymbol{\Phi}$ is unitary
\begin{equation}
    \boldsymbol{\Phi}\triangleq \boldsymbol{\Theta}_{2}\mathbf{F}\boldsymbol{\Theta}_{1},
\end{equation}
where $\boldsymbol{\Theta}_{1}$ and $\boldsymbol{\Theta}_{2}$ are rotational matrices defined as 
\begin{equation} \small
\boldsymbol{\Theta}_{1}=\mathrm{diag}\!\left(e^{-j2\pi c_{1}n^{2}}\right)_{n=0}^{N-1},
\quad
\boldsymbol{\Theta}_{2}=\mathrm{diag}\!\left(e^{-j2\pi c_{2}n^{2}}\right)_{n=0}^{N-1},
\end{equation}
where $c_1$ and $c_2$ are the AFDM design parameters, chosen as $
c_1=\frac{2k_{\max}+1}{2N},\text{ }c_2=\frac{1}{2N}$, 
with $k_{\max}\in\mathbb{Z}^{+}$ denoting the maximum normalized Doppler.\\
From \eqref{eq:I/O afdm} the effective channel of the AFDM is written as 
\begin{equation}
\mathbf{H}_{\mathrm{eff}} \triangleq \boldsymbol{\Phi}\,\mathbf{H}\,\boldsymbol{\Phi}^{H},
\end{equation}
with the entries of $\mathbf{H}_{\mathrm{eff}}$ given by
\begin{equation}
H^{\mathrm{af}}_{\mathrm{eff}}[m,n]
=\sum_{p=0}^{N-1}\sum_{q=0}^{N-1}\Phi[m,p]\;H[p,q]\;\Phi^{\ast}[q,n].
\label{eq:Heff_afdm_entry}
\end{equation}
Next, to proceed with \eqref{eq:Heff_afdm_entry} derivations, the product $\mathbf{D}\mathbf{P}$ from \eqref{eq:H_global_sum} is given for any ray $(c,r)$ as 
\begin{equation} \small
\big[\mathbf{D}\mathbf{P}\big][p,q]
=
\exp(j2\pi\frac{k_{c,r}^{\mathrm{tot}}}{N}p)\,
\frac{1}{N}\sum_{u=0}^{N-1}\exp\!\left(j2\pi\frac{u}{N}\big(p-q-l_{c}^{\mathrm{tot},i}\big)\right).
\label{eq:DP_entry_expand}
\end{equation}
Based on \eqref{eq:H_global_sum} and by substituting \eqref{eq:DP_entry_expand} in \eqref{eq:Heff_afdm_entry}, $H^{\mathrm{af}}_{\mathrm{eff}}$ becomes 
\begin{equation}
\begin{aligned}
&H^{\mathrm{af}}_{\mathrm{eff}}[m,n]
=\sum_{i=1}^{\mathcal{I}}\sum_{c=1}^{C_i}\sum_{r=1}^{R_{c,i}}
\frac{h_{c,r}^{(i)}}{N}
\sum_{p\in\mathcal{N}_i}\sum_{q=0}^{N-1}
\Phi[m,p]
 \\
&\times \exp\left(j2\pi\frac{k_{c,r}^{\mathrm{tot}}}{N}p\right)
\sum_{u=0}^{N-1}\exp\!\left(j2\pi\frac{u}{N}\big(p-q-l_{c}^{\mathrm{tot},i}\big)\right) \Phi^{\ast}[q,n].
\end{aligned}
\label{eq:Heff_afdm_mid1}
\end{equation}
Next, the order of summations  is exchanged and each variable dependent phase terms is separate, which gives
\begin{equation} \small
\begin{aligned}
&H^{\mathrm{af}}_{\mathrm{eff}}[m,n]
=\sum_{i=1}^{\mathcal{I}}\sum_{c=1}^{C_i}\sum_{r=1}^{R_{c,i}}
\frac{h_{c,r}^{(i)}}{N}
\sum_{u=0}^{N-1} \exp\!\left(-j2\pi\frac{u}{N}l_{c}^{\mathrm{tot},i}\right) \\
&
\sum_{p\in\mathcal{N}_i}\Phi[m,p]\;\exp\!\left(j2\pi\frac{k_{c,r}^{\mathrm{tot}}+u}{N}p\right)
\sum_{q=0}^{N-1}\Phi^{\ast}[q,n]\;\exp\!\left(-j2\pi\frac{u}{N}q\right).
\end{aligned}
\label{eq:Heff_afdm_mid2}
\end{equation}
Finally, the $\Phi[m,p]=\frac{1}{\sqrt{N}}e^{-j2\pi c_{2}m^{2}}\;e^{-j2\pi c_{1}p^{2}}\;e^{-j2\pi\frac{m}{N}p}$ kernel is substituted in \eqref{eq:Heff_afdm_mid2} which results in
\begin{equation}
\begin{aligned}
&H_{\mathrm{eff}}^{\mathrm{af}}[m,n]
=\sum_{i,c,r}\frac{h_{c,r}^{(i)}}{N^{2}}\;
e^{-j2\pi c_{2}m^{2}}\;e^{j2\pi c_{2}n^{2}}
\sum_{u=0}^{N-1} e^{-j2\pi \frac{u}{N}l_{c}^{\mathrm{tot},i}}\\
&\quad\times
\sum_{p\in\mathcal{N}_{i}} e^{-j2\pi c_{1}p^{2}}\;e^{j2\pi \frac{k_{c,r}^{\mathrm{tot}}+u-m}{N}p}
\sum_{q=0}^{N-1} e^{j2\pi c_{1}q^{2}}\;e^{j2\pi \frac{n-u}{N}q}.
\end{aligned}
\label{eq:Heff_afdm_final}
\end{equation}
From the \eqref{eq:Heff_afdm_final}, the Doppler spectrum appears via the ray index $r$; for a fixed cluster delay $l_c^{\mathrm{tot},i}$, the set $\{k_{c,r}^{\mathrm{tot}}\}$ induces multiple Doppler dependent phase slopes in the $p$ sum, so one delay cluster yields a superposition of shifted pulses in the affine domain, resulting in a broadening that widens with $R_{c,i}$. Fractional delay enters through the $u$ summation via $\exp(-j2\pi \tfrac{u}{N}l_c^{\mathrm{tot},i})$, when $l_c^{\mathrm{tot},i}\notin\mathbb{Z}$, the coupling of $u$ with the $p$ and $q$ dependent terms produces a leakage across affine indices. The non-WSSUS behavior appears through $p\in\mathcal{N}_i$ and region dependent $\{h_{c,r}^{(i)},l_c^{\mathrm{tot},i},k_{c,r}^{\mathrm{tot}},R_{c,i}\}$; windowing over $\mathcal{T}_i$ broadens the pulses as a function of the  birth and death MPCs \cite{gutierrez2017geometry}.
To scale down to the standard AFDM $\mathbf{H_{\mathrm{eff}}}$ \cite{b5}, a single stationarity region is assumed with $\mathcal{T}_0=\{0,\ldots,N-1\}$, one ray per cluster is retained $(R_{c,0}=1)$, and integer delay and Doppler shifts are imposed $l_c\in\mathbb{Z}$ and $k_c\in\mathbb{Z}$.\\
Under integer delay $l_c\in\mathbb{Z}$, the inner $q$ summation enforces $u = n$ via the DFT orthogonality identity $\sum_{q=0}^{N-1}e^{j2\pi(n-u)q/N}=N\,\delta[n-u]$, which collapses the $u$ summation to the single term $u=n$. The fractional delay operator then reduces to a circular shift, $P(l_c)[p,q]=\delta\!\big([p-q-l_c]_N\big)$, and the remaining
$p$ summation over the full frame yields the diagonal
AFDM structure. Simplifying the remaining phase terms leads to the conventional AFDM effective channel
\begin{equation}
\begin{aligned}
    H_{\mathrm{eff}}[m,n]
& =\sum_{c=1}^{C} h_{c}\,e^{-j2\pi c_{2}m^{2}}\,e^{j2\pi c_{2}n^{2}}
\; \\ &\times \delta\Big[n-\big[m-k_{c}+(2k_{max}+1)l_{c}\big]_N\Big],
\end{aligned}
\end{equation}
where $k_{\max} = \max |k_{c,r}^{\mathrm{tot}}|$.\\
\subsection{OFDM Waveform}
The received vector is obtained by applying a DFT matrix to the time domain observation, i.e.,
\begin{equation}
\mathbf{y}_{\mathrm{f}}=\mathbf{F}\,\mathbf{y}
=\mathbf{F}\,\mathbf{H}\,\mathbf{F}^{H}\mathbf{x}_{\mathrm{f}}+\mathbf{w}.
\end{equation}
Starting from $\mathbf{H}_{\mathrm{eff}}=\mathbf{F}\mathbf{H}\mathbf{F}^{H}$, the $(m,n)$th entry admits the element-wise expansion
\begin{equation}
H^{\mathrm{f}}_{\mathrm{eff}}[m,n]
=\sum_{p=0}^{N-1}\sum_{q=0}^{N-1}F[m,p]\;H[p,q]\;F^{\ast}[q,n].
\label{eq:Heff_ofdm_entry}
\end{equation}
Next, the piecewise channel structure in \eqref{eq:H_global_sum} is substituted, which restricts the receive-time summation to $p\in\mathcal{T}_{i}$, and the entry-wise product in \eqref{eq:DP_entry_expand} is used. This yields
\begin{equation}\small
\begin{aligned}
& H^{\mathrm{f}}_{\mathrm{eff}}[m,n]
=\sum_{i=1}^{\mathcal{I}}\sum_{c=1}^{C_i}\sum_{r=1}^{R_{c,i}}
\frac{h_{c,r}^{(i)}}{N}
\sum_{p\in\mathcal{N}_i}\sum_{q=0}^{N-1}
F[m,p]
 \\
&\times \exp\!\left(j2\pi\frac{k_{c,r}^{\mathrm{tot}}}{N}p\right)
\sum_{u=0}^{N-1}\exp\!\left(j2\pi\frac{u}{N}\big(p-q-l_{c}^{\mathrm{tot},i}\big)\right) F^{\ast}[q,n].
\end{aligned}
\label{eq:Heff_ofdm_mid1}
\end{equation}
Next, the order of summations is exchanged and the $u$ dependent phase terms are separated, which gives
\begin{equation} \small
\begin{aligned}
&H^{\mathrm{f}}_{\mathrm{eff}}[m,n]
=\sum_{i=1}^{\mathcal{I}}\sum_{c=1}^{C_i}\sum_{r=1}^{R_{c,i}}
\frac{h_{c,r}^{(i)}}{N}
\sum_{u=0}^{N-1} \exp\!\left(-j2\pi\frac{u}{N}l_{c}^{\mathrm{tot},i}\right) \\
&
\sum_{p\in\mathcal{N}_i}F[m,p]\;\exp\!\left(j2\pi\frac{k_{c,r}^{\mathrm{tot}}+u}{N}p\right) \sum_{q=0}^{N-1}F^{\ast}[q,n]\;\exp\!\left(-j2\pi\frac{u}{N}q\right).
\end{aligned}
\label{eq:Heff_ofdm_mid2}
\end{equation}
Finally, the DFT kernel entries $F[m,p]=\frac{1}{\sqrt{N}}e^{-j2\pi \frac{m}{N}p}$ and $F^{\ast}[q,n]=\frac{1}{\sqrt{N}}e^{j2\pi \frac{n}{N}q}$ are substituted into \eqref{eq:Heff_ofdm_mid2}. Collecting the phase terms with respect to $p$ and $q$ yields the final explicit OFDM effective channel expression
\begin{equation} \small
\begin{aligned}
H^{\mathrm{f}}_{\mathrm{eff}}[m,n]
&=\sum_{i,c,r}\frac{h_{c,r}^{(i)}}{N^{2}}
\sum_{u=0}^{N-1} e^{-j2\pi \frac{u}{N}l_{c}^{\mathrm{tot},i}}
\sum_{p\in\mathcal{N}_{i}} e^{j2\pi \frac{k_{c,r}^{\mathrm{tot}}+u-m}{N}p} \\ & \times
\sum_{q=0}^{N-1} e^{j2\pi \frac{n-u}{N}q}.
\end{aligned}
\label{eq:Heff_ofdm_final}
\end{equation}
From the OFDM effective channel expression, Doppler spectrum enters via the ray index $r$ through the $p$ sum term $\exp\!\big(j2\pi \tfrac{k_{c,r}^{\mathrm{tot}}+u-m}{N}p\big)$, yielding an inter carrier pulse leakage whose spread increases with $R_{c,i}$. Delay contributes through $\exp(-j2\pi \tfrac{u}{N}l_c^{\mathrm{tot},i})$ and reduces to a per subcarrier phase rotation under circular on grid conditions, thus ICI is driven primarily by Doppler and time variation over $\mathcal{T}_i$. The piecewise effect enters through $p\in\mathcal{N}_i$ and region dependent parameters, where truncation broadens inter-carrier coupling and coefficient refresh across $i$ yields region dependent responses.
To scale down to the standard OFDM setting, assume $\mathcal{T}_0=\{0,\ldots,N-1\}$, $R_{c,0}=1$, and $l_c,k_c\in\mathbb{Z}$. Then the $q$ sum enforces $u=n$ and the full-length $p$ sum yields a shifted diagonal form. Specifically, the downscaled OFDM effective channel is
\begin{equation}
H_{\mathrm{eff}}^{\mathrm{f}}[m,n]
=\sum_{c=1}^{C} h_{c}\,e^{-j2\pi \frac{n}{N}l_{c}}\,
\delta\Big[m-\big[n+k_{c}\big]_N\Big],
\end{equation}
which becomes diagonal in the quasi static case $k_c=0$, with diagonal entries $\sum_{c} h_{c}e^{-j2\pi \frac{n}{N}l_{c}}$.
\begin{table*}[t]
\centering
\caption{Representative dense-urban mobility operating points for $f_c=3.5$ GHz, $\tau_{\max}=0.5~\mu$s, $B\in\{20,100\}$ MHz, and  $\Delta f\in\{500~\mathrm{Hz},60~\mathrm{kHz}\}$ with normalized delay drift within one symbol quantified as $\Delta l_{\mathrm{sym}}\approx B\frac{v}{c}T$.}
\label{tab:motivation_regimes}
\renewcommand{\arraystretch}{1.12}
\setlength{\tabcolsep}{4.2pt}
\begin{tabular}{c c c c c c c c}
\hline
$v$ & $f_D$ & $l_{\max}$ & $l_{\max}$ & $k_{\max}$ & $k_{\max}$ & $\Delta l_{\mathrm{sym}}$ & $\Delta l_{\mathrm{sym}}$ \\
(km/h) & (Hz) & $B=20$ MHz & $B=100$ MHz & $\Delta f=500$ Hz & $\Delta f=60$ kHz & $\Delta f=500$ Hz & $\Delta f=60$ kHz \\
\hline
$30$  & $97.2$  & $10$ & $50$ & $0.1944$ & $0.0016$ & $(0.0011, 0.0056)$ & $(0.0000, 0.0000)$ \\
$120$ & $388.9$ & $10$ & $50$ & $0.7778$ & $0.0065$ & $(0.0044, 0.0222)$ & $(0.0000, 0.0002)$ \\
$300$ & $972.2$ & $10$ & $50$ & $1.9444$ & $0.0162$ & $(0.0111, 0.0556)$ & $(0.0001, 0.0005)$ \\
\hline
\end{tabular}
\end{table*}
\subsection{OTFS Waveform}
The OTFS domain received vector is obtained by applying the \ac{SFFT} analysis transform to the time domain observation. Following the Kronecker structured similarity transform in \cite{b19}, this is written as
\begin{equation}
\mathbf{y}_{\mathrm{dd}}=\mathbf{A}_{\mathrm{otfs}}\,\mathbf{y}
=\mathbf{A}_{\mathrm{otfs}}\,\mathbf{H}\,\mathbf{A}_{\mathrm{otfs}}^{H}\mathbf{x}_{\mathrm{dd}}+\mathbf{w},
\end{equation}
which motivates the OTFS effective channel matrix
\begin{equation}
\mathbf{H}^{\mathrm{dd}}_{\mathrm{eff}} \triangleq \mathbf{A}_{\mathrm{otfs}}\,\mathbf{H}\,\mathbf{A}_{\mathrm{otfs}}^{H}.
\end{equation}
Here, the OTFS kernel is written as $\mathbf{A}_{\mathrm{otfs}}\triangleq \mathbf{F}_{N'}\otimes \mathbf{I}_{M'}$, with $M'N'=N$.
Starting from $\mathbf{H}^{\mathrm{dd}}_{\mathrm{eff}}=\mathbf{A}_{\mathrm{otfs}}\mathbf{H}\mathbf{A}_{\mathrm{otfs}}^{H}$, the entry indexed by $(\mu,a)$ and $(\nu,b)$ admits the element-wise expansion
\begin{equation}
\begin{aligned}
    H^{\mathrm{dd}}_{\mathrm{eff}}\!\big[(\mu,a),(\nu,b)\big]
=\sum_{p=0}^{N-1}\sum_{q=0}^{N-1}&A_{\mathrm{otfs}}\!\big[(\mu,a),p\big]\;H[p,q]\;\\ &\times A_{\mathrm{otfs}}^{\ast}\!\big[q,(\nu,b)\big].
\end{aligned}
\label{eq:Heff_otfs_entry}
\end{equation}
Next, the index mapping $p=a+l M'$ and $q=b+l' M'$ is introduced, with $l,l'\in\{0,\ldots,N'-1\}$. Using the Kronecker structure of $\mathbf{A}_{\mathrm{otfs}}$, the kernel entries satisfy
\begin{equation}
\begin{aligned}
& A_{\mathrm{otfs}}\!\big[(\mu,a),a+l M'\big]=\frac{1}{\sqrt{N'}}e^{-j2\pi\frac{\mu l}{N'}},
\qquad \\ & 
A_{\mathrm{otfs}}^{\ast}\!\big[(\nu,b),b+l' M'\big]=\frac{1}{\sqrt{N'}}e^{+j2\pi\frac{\nu l'}{N'}}.
\end{aligned}
\label{eq:A_otfs_entries}
\end{equation}
Substituting \eqref{eq:A_otfs_entries} into \eqref{eq:Heff_otfs_entry} yields
\begin{equation}
\begin{aligned}
   H^{\mathrm{dd}}_{\mathrm{eff}}\!\big[(\mu,a),(\nu,b)\big]
=\frac{1}{N'}\sum_{l=0}^{N'-1}  & \sum_{l'=0}^{N'-1} 
e^{-j2\pi\frac{\mu l}{N'}}\;e^{+j2\pi\frac{\nu l'}{N'}}\; \\ &  
H\!\big[a+l M',\,b+l' M'\big].
\label{eq:Heff_otfs_mid1} 
\end{aligned}
\end{equation}
Next, the piecewise channel structure in \eqref{eq:H_global_sum} is substituted, which restricts the receive-time dependence to $a+l M'\in\mathcal{N}_i$, and the entry-wise product in \eqref{eq:DP_entry_expand} is used, yielding to \eqref{eq:Heff_otfs_final}.\\
From the OTFS effective channel expression, the Doppler spectrum appears through the ray index $r$: for a fixed cluster delay $l_c^{\mathrm{tot},i}$, the set $\{k_{c,r}^{\mathrm{tot}}\}$ induces multiple Doppler dependent phase slopes through the OTFS kernel, so one delay cluster yields a superposition of coupled delay-Doppler components on the $(M',N')$ grid, resulting in a broadening that increases with $R_{c,i}$. Fractional delay enters through the $u$ summation applied to $p-q-l_c^{\mathrm{tot},i}$; when $l_c^{\mathrm{tot},i}\notin\mathbb{Z}$, the coupling of this term with the index relation $p-q=(a-b)+M'(l-l')$ produces leakage across neighboring delay and Doppler bins. The non-WSSUS behavior appears through $a+lM'\in\mathcal{N}_i$ and region dependent $\{h_{c,r}^{(i)},l_c^{\mathrm{tot},i},k_{c,r}^{\mathrm{tot}},R_{c,i}\}$: windowing over $\mathcal{N}_i$ breaks the full Kronecker combining and broadens the response as a function of the birth and death of MPCs.\\
To scale down to the standard OTFS setting, assume $\mathcal{N}_0=\{0,\ldots,N-1\}$, $R_{c,0}=1$, and $l_c,k_c\in\mathbb{Z}$. Then the fractional delay kernel collapses to a circular shift and the Kronecker DFT yields the standard sparse mapping on the OTFS grid. The downscaled OTFS effective channel admits is give as
\begin{equation} \setcounter{equation}{45}
H_{\mathrm{eff}}^{\mathrm{dd}}\!\big[(\mu,a),(\nu,b)\big]
=\sum_{c=1}^{C} h_{c}\;
\delta\!\Big[\big[a-b-l_{c}\big]_{M'}\Big]\;
\delta\!\Big[\big[\mu-\nu-k_{c}\big]_{N'}\Big], 
\end{equation}
which corresponds to a 2 d circular convolution on the $(M',N')$ OTFS grid.
\subsection{DFT-s-OFDM Waveform}
The DFT-s-OFDM domain received vector is obtained by applying DFT de-spreading to the frequency domain observation on the allocated subcarriers. Assuming localized mapping onto a contiguous set $\mathcal{K}=\{k_{0},\ldots,k_{0}+N_{d}-1\}$, this is written as
\begin{equation}
\mathbf{y}_{\mathrm{DFT}}=\mathbf{F}_{N_{d}}\,\mathbf{y}_{\mathcal{K}}
=\mathbf{F}_{N_{d}}\,\mathbf{H}_{\mathcal{K}}\,\mathbf{F}_{N_{d}}^{H}\mathbf{x}_{\mathrm{DFT}}+\mathbf{w},
\end{equation}
which motivates the DFT-s-OFDM effective channel matrix
\begin{equation}
\mathbf{H}_{\mathrm{eff}} \triangleq \mathbf{F}_{N_{d}}\,\mathbf{H}_{\mathcal{K}}\,\mathbf{F}_{N_{d}}^{H},
\end{equation}
where $\mathbf{H}_{\mathcal{K}}\in\mathbb{C}^{N_{d}\times N_{d}}$ denotes the submatrix of $\mathbf{F}\mathbf{H}\mathbf{F}^{H}$ restricted to the allocated subcarriers, i.e.,
\begin{equation}
\big[\mathbf{H}_{\mathcal{K}}\big]_{a,b}\triangleq \big[\mathbf{F}\mathbf{H}\mathbf{F}^{H}\big]_{k_{0}+a,\,k_{0}+b},
\qquad a,b=0,\ldots,N_{d}-1.
\label{eq:Hk_def}
\end{equation}

Starting from $\mathbf{H}_{\mathrm{eff}}=\mathbf{F}_{N_{d}}\mathbf{H}_{\mathcal{K}}\mathbf{F}_{N_{d}}^{H}$, the $(m,n)$th entry admits the element-wise expansion
\begin{equation}
H^{\mathrm{DFT}}_{\mathrm{eff}}[m,n]
=\sum_{a=0}^{N_{d}-1}\sum_{b=0}^{N_{d}-1}F_{N_{d}}[m,a]\;\big[\mathbf{H}_{\mathcal{K}}\big]_{a,b}\;F_{N_{d}}^{\ast}[b,n].
\label{eq:Heff_sc_entry}
\end{equation}
Substituting \eqref{eq:Hk_def} into \eqref{eq:Heff_sc_entry} yields
\begin{equation}
\begin{aligned}
    H^{\mathrm{DFT}}_{\mathrm{eff}}[m,n]
=\sum_{a=0}^{N_{d}-1}\sum_{b=0}^{N_{d}-1}& F_{N_{d}}[m,a]\;\big[\mathbf{F}\mathbf{H}\mathbf{F}^{H}\big]_{k_{0}+a,\,k_{0}+b}\; \\ & \times F_{N_{d}}^{\ast}[b,n].
\end{aligned}
\label{eq:Heff_sc_mid0}
\end{equation}
Using the element-wise representation of $\mathbf{F}\mathbf{H}\mathbf{F}^{H}$ and the entry-wise product $\big[\mathbf{D}(\cdot)\mathbf{P}(\cdot)\big]_{p,q}$ in \eqref{eq:DP_entry_expand}, and substituting the piecewise channel structure leads to \eqref{eq:Heff_sc_final}.
\par From the DFT-s-OFDM effective channel expression, the Doppler spectrum appears through the ray index $r$: for a fixed cluster delay $l_c^{\mathrm{tot},i}$, the set $\{k_{c,r}^{\mathrm{tot}}\}$ induces multiple Doppler dependent phase slopes in the $p$ summation, so one delay cluster yields a superposition of frequency couplings that, after despreading, results in a broadening across the DFT spread data indices. Fractional delay enters through the $u$ summation via $\exp(-j2\pi \tfrac{u}{N}l_c^{\mathrm{tot},i})$; when $l_c^{\mathrm{tot},i}\notin\mathbb{Z}$, the coupling of $u$ with the allocated tones and the despreading sums prevents a pure per-tone phase representation and produces leakage across the data domain. The non-WSSUS behavior appears through $p\in\mathcal{N}_i$ and region dependent $\{h_{c,r}^{(i)},l_c^{\mathrm{tot},i},k_{c,r}^{\mathrm{tot}},R_{c,i}\}$: windowing over $\mathcal{N}_i$ broadens the coupling after despreading, while coefficient refresh and birth and death of MPCs yield region dependent DFT-s-OFDM responses.\\
 \begin{figure}[!t]
    \centering
  \includegraphics[width=\columnwidth]{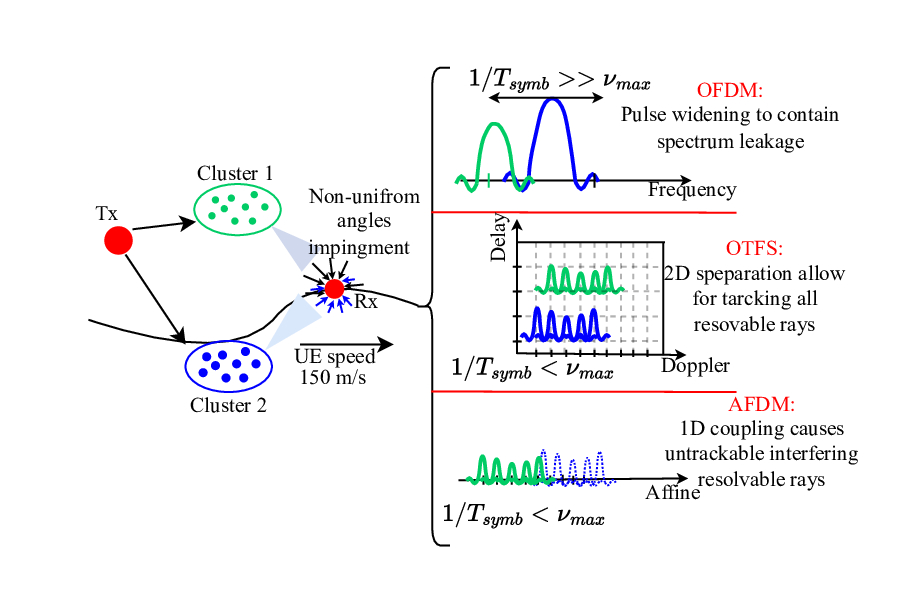}
    \caption{Pulse shape interpretation for Doppler spectrum in different waveform domains under non-uniform angular impingement.}
    \label{fig: Doppler_spec}
\end{figure}
To scale down to the standard DFT-s-OFDM setting, assume $\mathcal{N}_0=\{0,\ldots,N-1\}$, $R_{c,0}=1$, and $l_c,k_c\in\mathbb{Z}$. Then the full-length sums enforce the standard discrete selections on the allocated tones and de-spreading yields the standard DFT s-OFDM effective channel
\begin{equation} \small \setcounter{equation}{52}
H_{\mathrm{eff}}^{\mathrm{DFT}}[m,n]
=\sum_{c=1}^{C} h_{c}\,
\delta\!\Big[m-\big[n+k_{c}\big]_{N_d}\Big]\,
\exp\!\left(-j2\pi\frac{k_0+n}{N}l_c\right),
\end{equation}
where the delay dependent phase is evaluated on the allocated subcarriers and reduces to a constant phase across the data domain indices when localized mapping is used and $k_c=0$. \\
Fig. \ref{fig: CH_eff_representation} depicts the visualization of the derived effective channel equations in \eqref{eq:Heff_ofdm_final}, \eqref{eq:Heff_afdm_final}, \eqref{eq:Heff_sc_final}, and \eqref{eq:Heff_otfs_final}. The first row shows the downscaled version of each $\mathbf{H_{eff}}$, where both AFDM and OTFS representations indicate sparse $\mathbf{H_{eff}}$ with resolvable diagonals that lead to diversity gains. However, upon projection to the proposed models, AFDM and OTFS representations indicate a huge power interference across both domains, whereas an OFDM representation shows a comparatively less interference.  
\begin{figure*}[t]
    \centering
    \includegraphics[width=\textwidth, height=0.20\textheight]{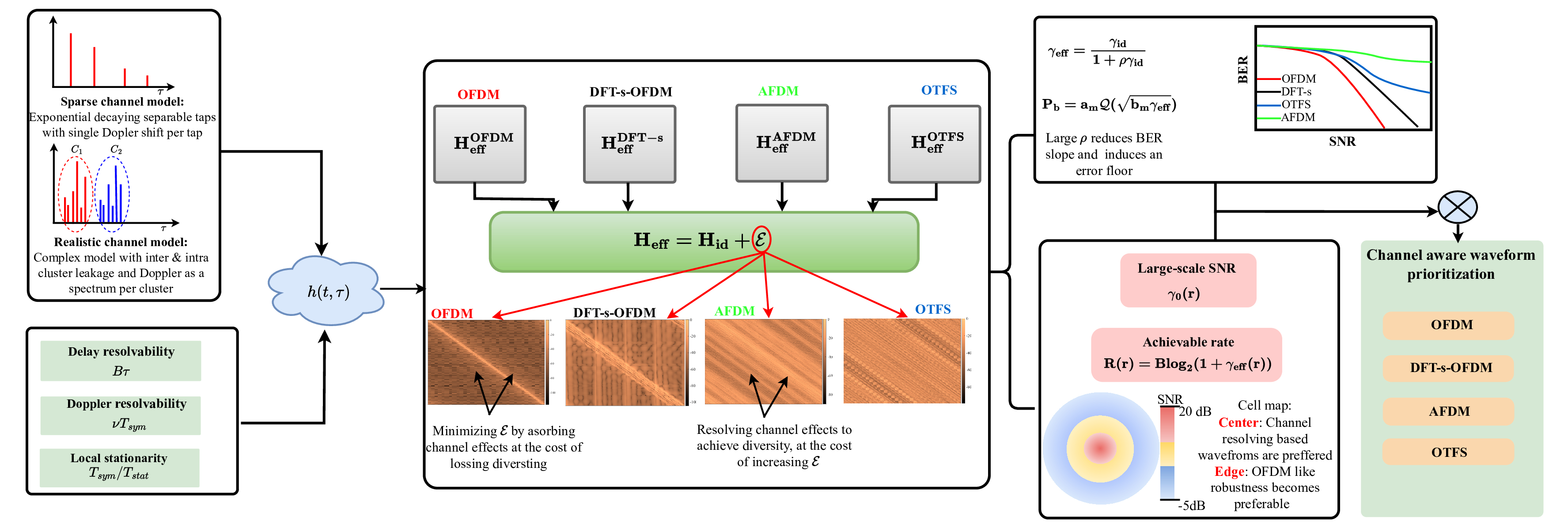}
    \caption{Block diagram illustrating the proposed channel dependent adaptive waveform prioritization framework.}
    \label{fig: block_diag_pro}
\end{figure*}
\par \textbf{Proof}. See Appendix for Section III derivations.
\section{Channel Aware Waveform Selection}
The effective channels in \eqref{eq:Heff_afdm_final}, \eqref{eq:Heff_ofdm_final}, \eqref{eq:Heff_otfs_final}, and \eqref{eq:Heff_sc_final} show that the same physical channel is mapped into different interference structures depending on the processing domain. Therefore, waveform selection should be based on how each waveform absorbs or resolves the channel effects through its corresponding effective channel.
\subsection{Effective Channel Leakage}
For bandwidth $B$ and a symbol duration $T$, the nominal resolutions are
$\Delta \tau_{\mathrm{res}}=\frac{1}{B},
\text{ }
\Delta \nu_{\mathrm{res}}=\frac{1}{T}.$ A delay pair is resolvable only if $|\tau_c-\tau_{c'}|\gtrsim \Delta \tau_{\mathrm{res}}$, whereas a Doppler pair is resolvable only if $|\nu_{c,r}-\nu_{c',r'}|\gtrsim \Delta \nu_{\mathrm{res}}$. Hence, delay resolvability requires large $B$, while Doppler resolvability requires large $T$. However, the same $T$ must remain within a stationarity region $T \leq T_{\mathrm{stat}}$. Thus, sparse domain operation is meaningful only when $\frac{1}{\Omega_{\nu}}\lesssim T \leq T_{\mathrm{stat}}$, where
\begin{equation}
\Omega_{\nu} \triangleq \max_{c,r,r'}\!\left|\nu_{c,r}-\nu_{c,r'}\right|
= 2f_D\,\sin\!\left(\tfrac{\Delta\theta_c}{2}\right),
\label{eq:omega_nu}
\end{equation}
denotes the effective Doppler spread width. If previous condition is violated, the channel cannot be both Doppler resolved and locally stationary, and the region dependent parameters $\{h_{c,r}^{(i)},l_c^{\mathrm{tot},i},R_{c,i}\}$ varies over each $\mathcal{N}_i$. let following normalized indicator defined as 
\begin{equation}
\chi_{\tau}\triangleq B\tau_{\max},
\qquad
\chi_{\nu}\triangleq f_D T,
\qquad
\chi_{\mathrm{stat}}\triangleq \frac{T}{T_{\mathrm{stat}}}.
\label{eq:chi_metrics}
\end{equation}
Where $\chi_{\tau}\gtrsim 1$ indicates delay resolvability, $\chi_{\nu}\gtrsim 1$ indicates Doppler resolvability, and $\chi_{\mathrm{stat}}\lesssim 1$ is required for local stationarity. Table~\ref{tab:motivation_regimes} shows that minimizing $\Delta f$, which increases $T$ to resolve Doppler leads to augmenting the normalized delay drift $\Delta l_{\mathrm{sym}}$ up to the order of fraction, whereas it diminishes when increasing $\Delta f$. Consequently, the waveforms that absorb channel effects, like OFDM in   \eqref{eq:Heff_ofdm_final} and DFT-s-OFDM in  \eqref{eq:Heff_sc_final}  minimize interference by shrinking $T$, while  \eqref{eq:Heff_afdm_final} and \eqref{eq:Heff_otfs_final} undergoes unavoidable interference due to $\Delta l_{\mathrm{sym}}$ through $T$ enlargement. For all waveforms, let the effective channel be written as
\begin{equation}
\mathbf{H}_{\mathrm{eff}}=\mathbf{H}_{\mathrm{id}}+\mathcal{E},
\label{eq:Heff_decomposition}
\end{equation}
where $\mathbf{H}_{\mathrm{id}}$ denotes the ideal downscaled effective channel and $\mathcal{E}$ collects the residual interference induced by fractional delay, fractional Doppler, Doppler-spectrum spreading, and segmentation leakage. The corresponding received symbol is modeled as
\begin{equation}
\mathbf{y}=\mathbf{H}_{\mathrm{id}}\mathbf{x}+\mathcal{E}\mathbf{x}+\mathbf{w},
\label{eq:effective_symbol_model}
\end{equation}
let
\begin{equation}
\gamma_{\mathrm{sig}}=\mathbb{E}\!\left[\|\mathbf{H}_{\mathrm{id}}\mathbf{x}\|_2^2\right],
\qquad
\gamma_{\mathrm{leak}}=\mathbb{E}\!\left[\|\mathcal{E}\mathbf{x}\|_2^2\right].
\label{eq:Psig_Pleak}
\end{equation}
If $\widehat{\mathbf{H}}_{\mathrm{eff}}$ denotes the effective channel reconstructed from \ac{CE}, the reconstruction mismatch is defined by
\begin{equation}
\gamma_{\mathrm{mm}}
=
\left\|
\mathbf{H}_{\mathrm{eff}}-\widehat{\mathbf{H}}_{\mathrm{eff}}
\right\|_F^2,
\label{eq:mismatch}
\end{equation}
and its symbol-domain contribution is denoted by $\gamma_{\mathrm{mm}}$. Hence, the effective SINR becomes
\begin{equation}
\gamma_{\mathrm{eff}}
=
\frac{\gamma_{\mathrm{sig}}}
{\sigma^2+\gamma_{\mathrm{leak}}+\gamma_{\mathrm{mm}}},
\label{eq:gamma_eff}
\end{equation}
with
\begin{equation}
\gamma_{\mathrm{id}}=\frac{\gamma_{\mathrm{sig}}}{\sigma^2},
\qquad
\rho = \frac{\gamma_{\mathrm{leak}}+\gamma_{\mathrm{mm}}}{\gamma_{\mathrm{sig}}},
\label{eq:rho}
\end{equation}
By substituting in \eqref{eq:gamma_eff}, the effective SINR becomes 
\begin{equation}
\gamma_{\mathrm{eff}}
=
\frac{\gamma_{\mathrm{id}}}{1+\rho\,\gamma_{\mathrm{id}}}.
\label{eq:gamma_eff_alt}
\end{equation}
Therefore, waveform-dependent BER differences follow directly from the corresponding $\mathbf{H}_{\mathrm{eff}}$, and thus $\rho$ is waveform dependent. If $\rho$ does not vanish with SNR, then
\begin{equation}
\lim_{\gamma_{\mathrm{id}}\rightarrow\infty}\gamma_{\mathrm{eff}}=\frac{1}{\rho},
\label{eq:sinr_floor}
\end{equation}
which induces a BER floor that is dependent on waveform properties. For Gray-mapped square $M$-QAM \cite{najafizadeh2006ber},
\begin{equation}
P_b \approx a_M Q\!\left(\sqrt{b_M\gamma_{\mathrm{eff}}}\right),
\label{eq:BER_gammaeff}
\end{equation}
where $a_M$ and $b_M$ are modulation dependent variables \cite{najafizadeh2006ber}. A major contributor $\rho$ through $\gamma_{\mathrm{mm}}$ is the ambiguity of distinguishing between a single Doppler shift with fractional spread or a Doppler spectrum spread. For \ac{OFDM}, Doppler spreading regardless of its source is absorbed with larger $\Delta f$ and through pulse shaping. For \ac{OTFS}, the ambiguity is partially resolvable by following the pulse distortion across the Doppler axis, benefiting from the 2D structure of OTFS that decouples delays and Doppler shifts. However, in \ac{AFDM}, the distinction becomes harder as pulses in the affine domain are already distorted by only a single Doppler shift \cite{b36}. Fig. \ref{fig: Doppler_spec} illustrates the Doppler spread spectrum behavior across different waveform.
\subsection{SNR Distribution Waveform Selection}
The previous subsection characterizes the small-scale behavior through $\mathbf{H}_{\mathrm{eff}}$, $\mathcal{E}$, and the resulting leakage ratio $\rho$. To extend the analysis to cell-level operation, let $\gamma_0(\mathbf{r})$ denote the received \ac{SINR} at cell position $\mathbf{r}$ due to large-scale propagation. For the same $\gamma_0(\mathbf{r})$, each waveform yields a different usable \ac{SINR} because its corresponding $\mathbf{H}_{\mathrm{eff}}$ induces a different leakage level. Hence, for each waveform candidate,
\begin{equation}
\gamma_{\mathrm{eff}}(\mathbf{r})
=
\frac{\gamma_0(\mathbf{r})}
{1+\rho\,\gamma_0(\mathbf{r})},
\label{eq:gammaeff_r}
\end{equation}
where $\rho$ is computed from the corresponding $\mathbf{H}_{\mathrm{eff}}$. Accordingly, the achievable rate at position $\mathbf{r}$ becomes
\begin{equation}
R(\mathbf{r})
=
B\log_2\!\big(1+\gamma_{\mathrm{eff}}(\mathbf{r})\big),
\label{eq:rate_r_cont}
\end{equation}
or, under discrete modulation adaptation,
\begin{equation}
R(\mathbf{r})
=
B\log_2 M^\star(\mathbf{r}),
\label{eq:rate_r_disc}
\end{equation}
where $M^\star(\mathbf{r})$ is the highest modulation order satisfying the target \ac{BER} at $\gamma_{\mathrm{eff}}(\mathbf{r})$. Therefore, even at the same cell position, different waveforms provide different rates since the same received \ac{SINR} is projected differently through their corresponding effective channels. In addition, \ac{PAPR} affects the usable \ac{SINR}, since waveforms with larger peaks become more sensitive to transmitter-side power limitations. Hence, beyond the leakage behavior induced by $\mathbf{H}_{\mathrm{eff}}$, the final operating point is also influenced by the waveform-dependent \ac{PAPR}.\\
Accordingly, Sections~IV-A and IV-B establish the proposed selection rule from two complementary levels. First, the small-scale analysis shows how each waveform transforms the same physical channel into a different $\mathbf{H}_{\mathrm{eff}}$, leading to different leakage and \ac{BER} behavior. Then, the cell-level view shows how the received \ac{SNR} at a given position is converted into a different effective \ac{SINR} and achievable rate for each waveform. Therefore, the proposed scheme is not a static waveform decision, but a channel-aware waveform selection rule driven by the structure of $\mathbf{H}_{\mathrm{eff}}$ and the resulting effective performance under the active regime. Fig. \ref{fig: block_diag_pro} illustrates a summary for the proposed channel aware waveform prioritization framework. 
\section{Simulation Results}
\begin{figure}[t]
    \centering
  \includegraphics[width=\columnwidth]{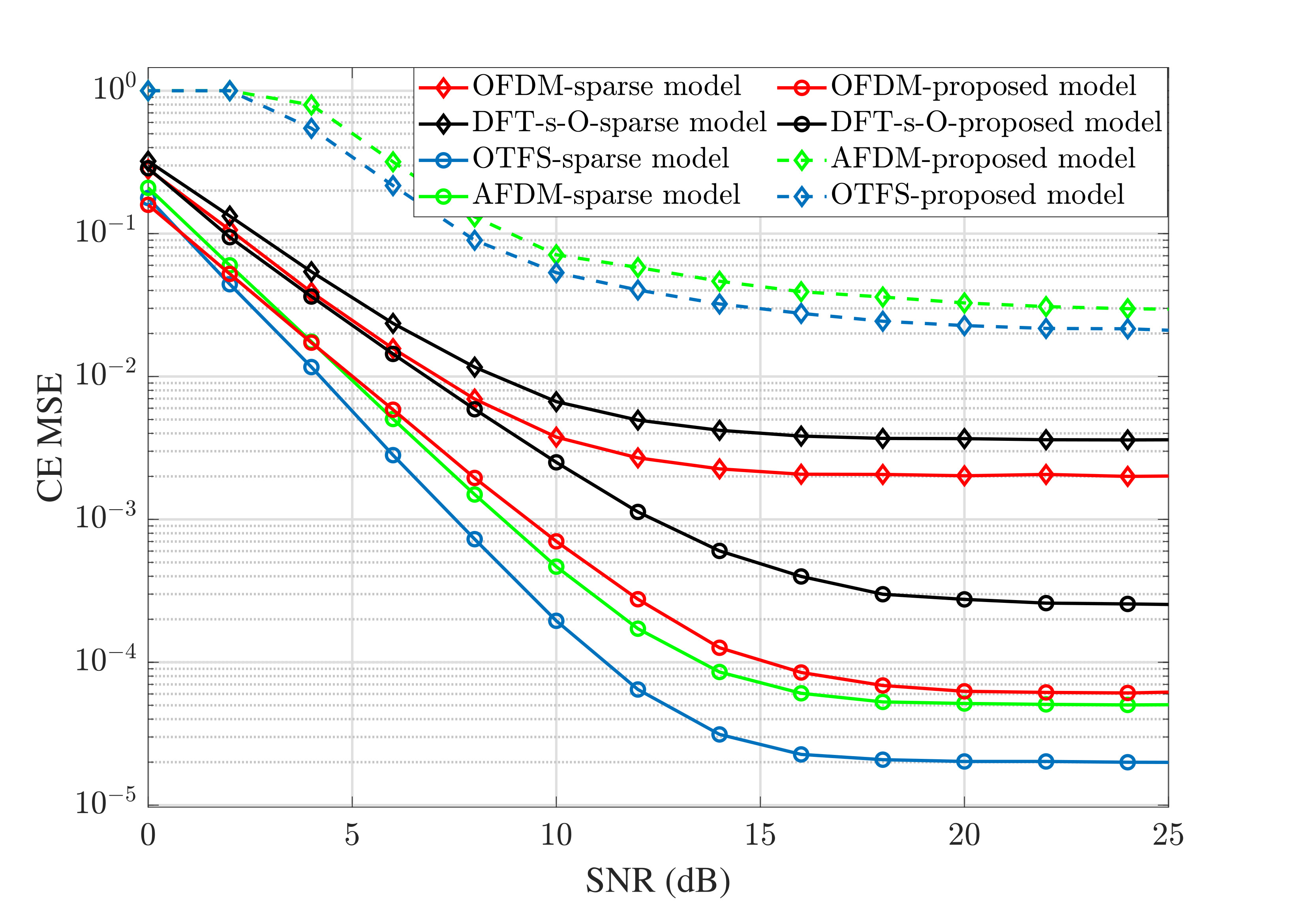}
    \caption{CE NMSE versus SNR under sparse and proposed channel models.}
    \label{fig:CE_S_P}
\end{figure}

This section evaluates the performance of \ac{OFDM}, \ac{DFT} s-\ac{OFDM}, \ac{AFDM}, and \ac{OTFS} under the sparse channel regime and the proposed channel regime, while validating the criteria for channel dependent waveform selection. In the simulations, Gray-mapped $M$-ary \ac{QAM} is used for data modulation over each waveform effective channel $\mathbf{H}_{\mathrm{eff}}$ derived in Section~III. The channel parameters $C,R,h_{c,r}, \nu_{c,r},\tau_{c} $ are generated based on the latest 3GPP release $18$ CDL-A measurements for \ac{NLoS} case, with carrier frequency $f_c=3.5$~GHz, maximum excess delay $\tau_{\max}=0.5~\mu$s, bandwidths $B\in\{20,100\}$~MHz, subcarrier spacings $\Delta f\in\{500~\mathrm{Hz},1~\mathrm{kHz},15~\mathrm{kHz},60~\mathrm{kHz}\}$, and user speeds up to $\{300\}$~km/h. Unless otherwise stated, all \ac{BER} results are obtained using \ac{MMSE} and \ac{MRC} equalization, and all curves are averaged over independent Monte Carlo channel realizations.\\
Fig.~\ref{fig:CE_S_P} illustrates the \ac{CE} versus SNR under both sparse and proposed channel models, where channel parameters are extracted in each waveform specific domain and mapped to reconstruct $\mathbf{H_{eff}}$. For the sparse channel model, \ac{CE} in OTFS and AFDM is based on the \ac{EPA} technique, which searches for channel parameters $h, \text{ } l, \text{ } k$ using a domain indices thresholding decision, allowing for a near-perfect reconstruction of $\mathbf{H_{eff}}$ susceptible only to  noise and \ac{MPCs} fractional leakage that can be overcome with pilots' power or grid size \cite{b5}. This reflects on OTFS and AFDM achieving the lowest NMSE under a sparse channel model. Whereas for OFDM and DFT s-OFDM, as \ac{CE} is based on phase tracking through pilot interpolation, the channel parameters are not independently estimated but rather their total effect, where the Doppler effect creates an error floor that pushes NMSE to higher values even under a sparse channel model. Furthermore, for the proposed channel model, the same \ac{CE} techniques are kept for all waveforms; however, the induced interference magnifies significantly between waveforms. For OFDM and DFT s-OFDM, as the total channel effect is estimated, the interference is absorbed by widening $\Delta f$, which results in preserving the WSSUS region, leading to a minimal degradation of around $1$ order of fraction only from Doppler spectrum leakage mainly. Meanwhile, AFDM and OTFS rely on \ac{MPCs} resolvability; both become susceptible to sinc leakage resulting from symbol segmentation due to breaking the WSSUS region, accumulated modeling errors of the $h_{c,r} \text{ and } l_c$, along with the ambiguity to distinguish Doppler spectrum from fractional spread. All these lead to an NMSE degradation of $3$ orders of fraction.
\begin{figure}[t]
    \centering
  \includegraphics[width=\columnwidth]{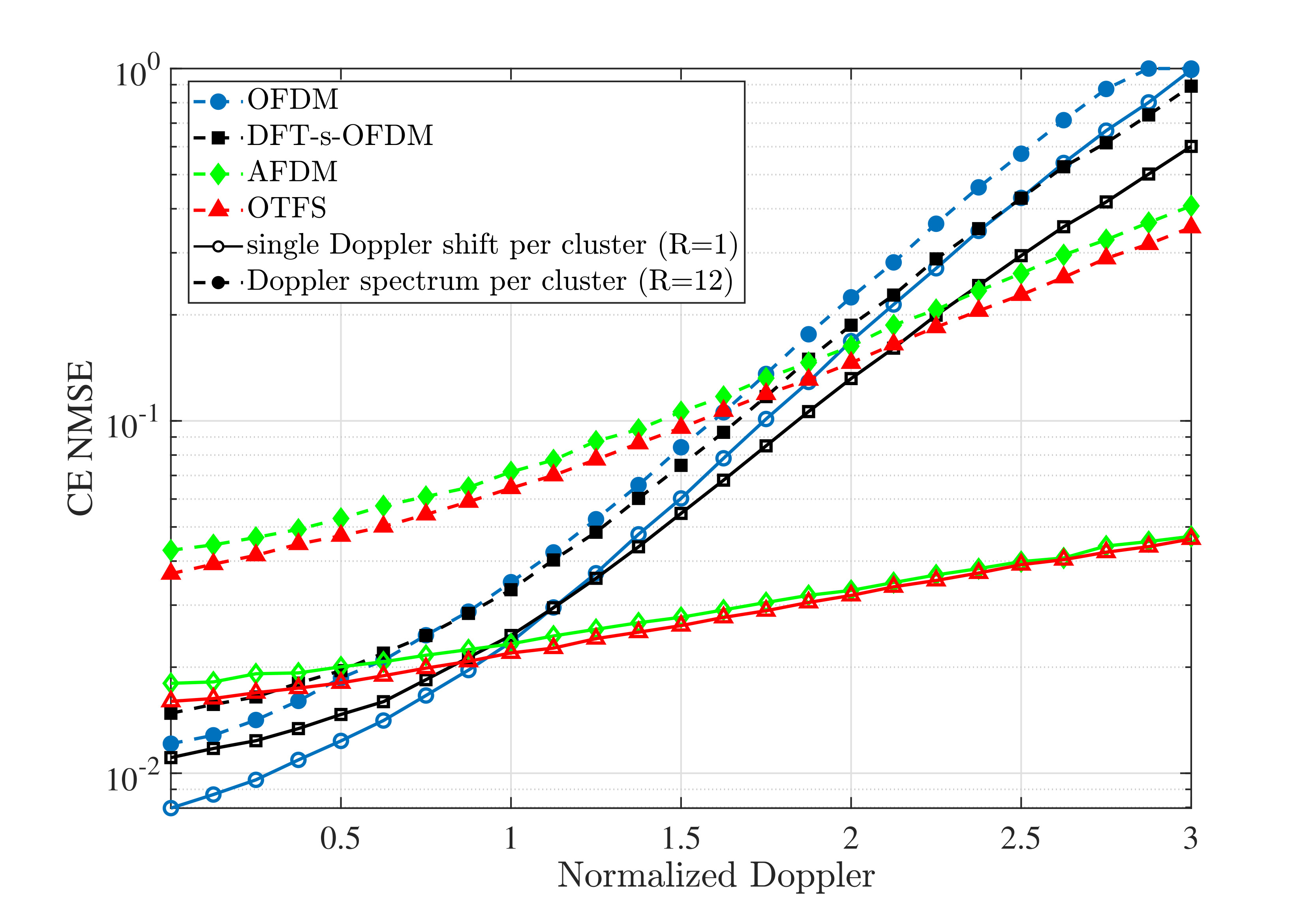}
    \caption{CE NMSE versus normalized Doppler under the proposed model for a single Doppler shift per cluster $(R=1)$ and a Doppler spectrum per cluster $(R=12)$.}
    \label{fig:CE_Doppler}
\end{figure}

\begin{figure}[t]
    \centering
  \includegraphics[width=\columnwidth]{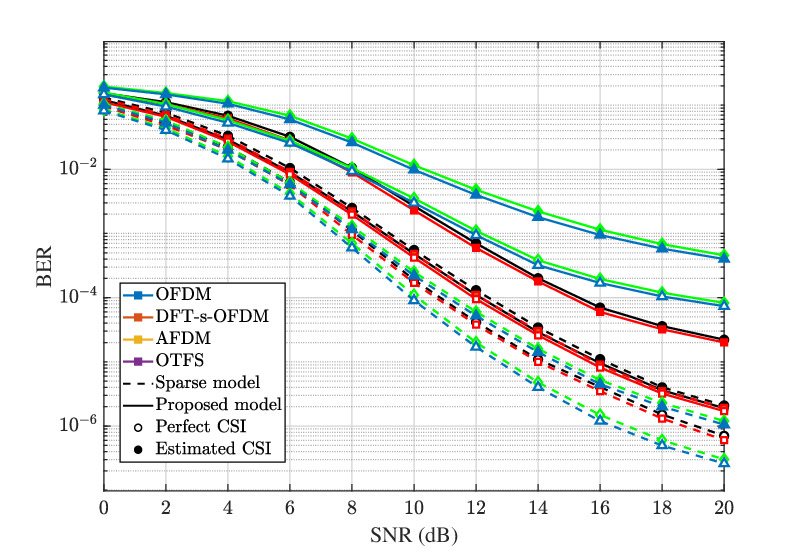}
    \caption{BER versus SNR under sparse and proposed channel models with perfect and estimated channel.}
    \label{fig:BER_SS}
\end{figure}

Fig.~\ref{fig:CE_Doppler} depicts the \ac{CE} accuracy as a function of the normalized Doppler index under the proposed channel model, while fixing $C$ to $12$ and comparing two cases: a single Doppler shift per cluster $R=1$ and a Doppler spectrum formed by multiple rays inside each cluster. For \ac{OFDM} and DFT s-OFDM, the \ac{NMSE} rises steadily with the normalized Doppler, as resolving Doppler creates high error floors when tracking phase variation through the \ac{CE} process, giving nearly similar degradation for both single Doppler shifts or Doppler spectra. Whereas, for \ac{AFDM} and \ac{OTFS}, both show a nearly stable \ac{CE} performance for the single shift case. However, when a Doppler spectrum is introduced inside each cluster, a huge degradation is observed as an ambiguity rises on modeling whether the resulting shift is due to fractional spread of a single MPC or Doppler spread spectrum. Although Doppler spectrum can be tracked in OTFS through pulse shape tracking, AFDM fails, as the 1D structure pushed the pulse to become nonuniform under a rich scattering channel \cite{b36}. Fig.~\ref{fig:CE_Doppler} indicates that under a practical model, OFDM achieves better channel tracking at low resolvable Doppler, which is its optimum point, compared to the optimum point of \ac{OTFS} and AFDM within the resolvable Doppler. \\
Fig .~\ref{fig:BER_SS} presents the raw \ac{BER} comparison under sparse and proposed channel models for both perfect and estimated channel information using \ac{MMSE} equalization for OFDM and DFTs-OFDM, while \ac{MRC} is used for AFDM and OTFS; the latter is used to ensure the convergence of OFDM and DFT s-OFDM BER curves to the optimum BER levels compared to \ac{LS}. Attributing to the ability to resolve MPCs with minimum correlation and coherent combining using \ac{MRC}, AFDM, and OTFS achieves the optimum BER levels under the sparse channel model compared to the other scenarios. However, when AFDM and OTFS are tested under the proposed channel model, the propagating \ac{CE} errors lead to unreliable BER even at high SNR values. Moreover, even under perfect channel information, AFDM and OTFS showcase a degraded BER due to the unhandled leakage arising from the break of the WSSUS region, which augments the correlation among MPCs.  For the OFDM and DFT s-OFDM, BER follows similar trend to CE curves. The degradation when assessed under the proposed model is absorbed by adjusting the $\Delta f$, allowing for maintaining within a reliable BER level at slightly higher SNR.\\
\begin{figure}[t]
    \centering
  \includegraphics[width=\columnwidth]{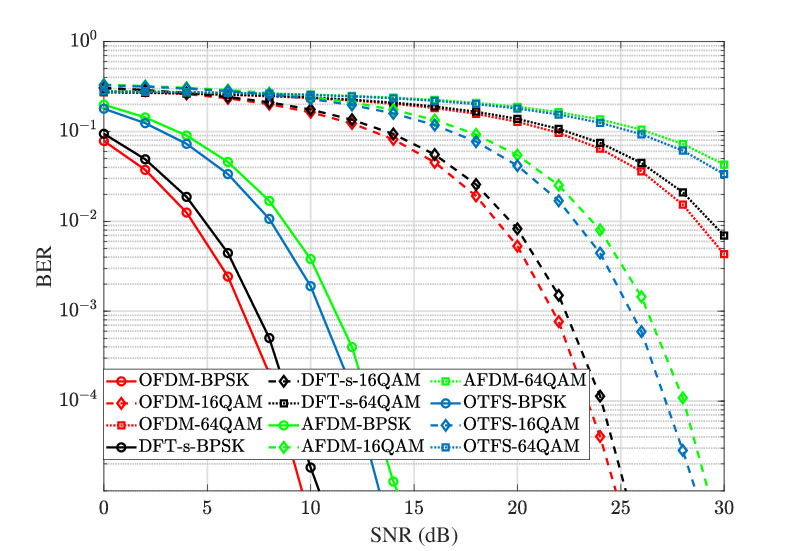}
    \caption{BER versus SNR under the proposed channel model for different modulation orders.}
    \label{fig:BER_modd}
\end{figure}
Fig.~\ref{fig:BER_modd} illustrates the \ac{BER} versus \ac{SNR} under the proposed channel model for different modulation orders. As expected, increasing the modulation order from low-order signaling to $16$-QAM and $64$-QAM shifts all curves to the right, since a larger constellation requires a higher effective \ac{SNR} to maintain reliable decision regions. It is observed that \ac{AFDM} and \ac{OTFS} suffer a noticeably larger penalty than \ac{OFDM} and \ac{DFT} s-\ac{OFDM}, with an \ac{SNR} degradation of about $5$ dB already appearing at low modulation order. This observation is particularly important because practical 5G NR and anticipated \ac{6G} frame structures rely mainly on low-order modulation for reference and control signaling. Therefore, the larger \ac{BER} penalty of \ac{AFDM} and \ac{OTFS} directly reduces fairness across the cell, since reliable low-order signaling cannot be sustained over the same coverage region. This trend becomes severe at higher orders, where \ac{AFDM} and \ac{OTFS} require stronger channel coding at moderate \ac{SNR} to reach the same reliability target. Under a fair comparison with \ac{OFDM} using similar coding, \ac{OFDM} remains superior under the proposed channel regime.\\
\begin{figure}[t]
    \centering
    \begin{subfigure}[b]{0.30\textwidth}
        \centering
        \includegraphics[width=\columnwidth]{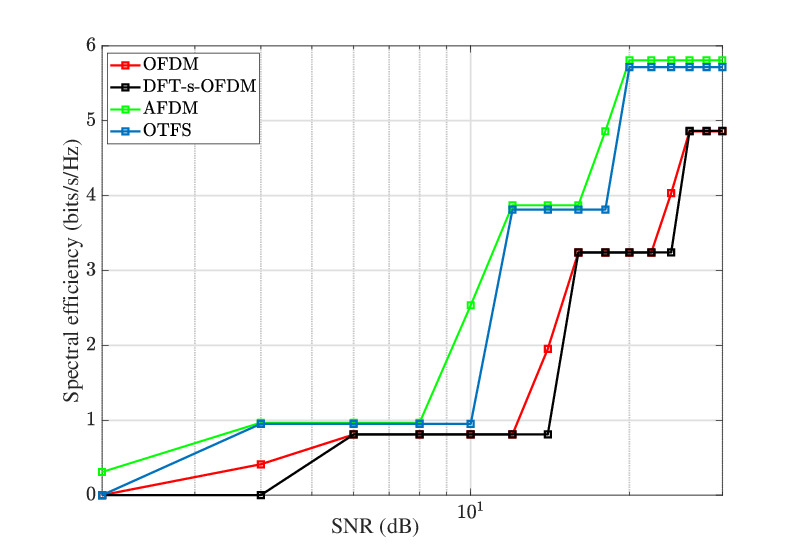}
        \caption{}
        \label{}
    \end{subfigure}\hfill
    \begin{subfigure}[b]{0.30\textwidth}
        \centering
        \includegraphics[width=\columnwidth]{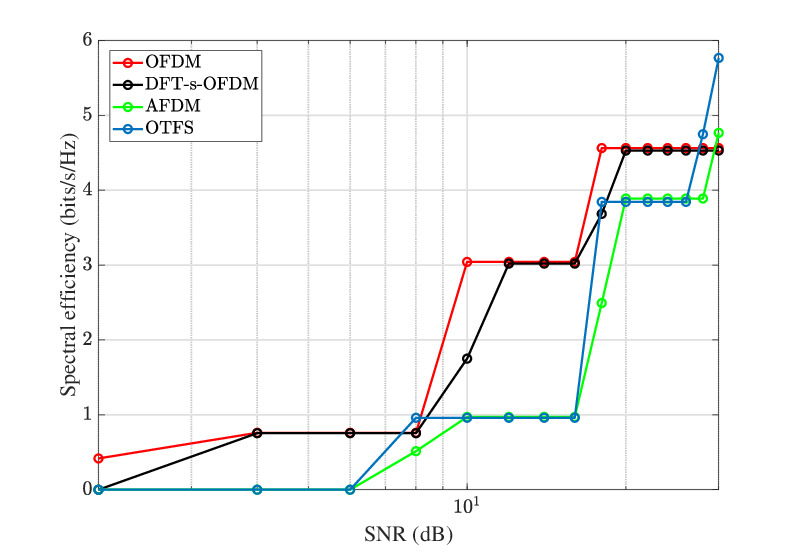}
        \caption{}
        \label{fig:SE_comp}
    \end{subfigure}
    \caption{Spectral efficiency with fair resource mapping bandwidth and time allocation: (a) sparse model and (b) proposed model.}
    \label{fig:SE_comp}
\end{figure}
Fig.~\ref{fig:SE_comp} evaluates the spectral efficiency under exact resource accounting for both sparse and proposed regimes. The effective spectral efficiency is computed as $\mathrm{SE}(\gamma)=\eta\,\log_2\!\big(M^\star(\gamma)\big)$, where $\eta=N_{\mathrm{data}}/N_{\mathrm{tot}}$ denotes the payload fraction after excluding \ac{CP}, pilot, and guard overhead, and $M^\star(\gamma)$ is the highest modulation order satisfying the target \ac{BER} at \ac{SNR} $\gamma$. Hence, the comparison reflects not only reliability, but also the actual usable data portion within the same time-bandwidth resource. For \ac{OFDM} and \ac{DFT} s-\ac{OFDM}, achieving reliable \ac{BER} under mobility requires increasing $\Delta f$, which shortens the symbol duration and improves robustness at the cost of increasing the \ac{CP}-to-symbol ratio. In contrast, \ac{AFDM} and \ac{OTFS} require a larger observation support in time, together with pilot protection and guards, in order to resolve more channel parameters. Under the sparse model, \ac{AFDM} and \ac{OTFS} still achieve higher spectral efficiency once the \ac{SNR} is sufficiently large, since their sparse-domain operation remains effective and the additional overhead is compensated by earlier access to higher modulation orders. Under the proposed model, however, the trend shifts in favor of \ac{OFDM} and \ac{DFT} s-\ac{OFDM}, since the extra spreading and guard cost of \ac{AFDM} and \ac{OTFS} is no longer justified by a reliable sparse reconstruction. Therefore, the spectral-efficiency comparison confirms that the preferable waveform is the one that achieves the target \ac{BER} with the smallest total resource cost under the active regime.\\
   \begin{figure}[!t]
    \centering
  \includegraphics[width=\columnwidth]{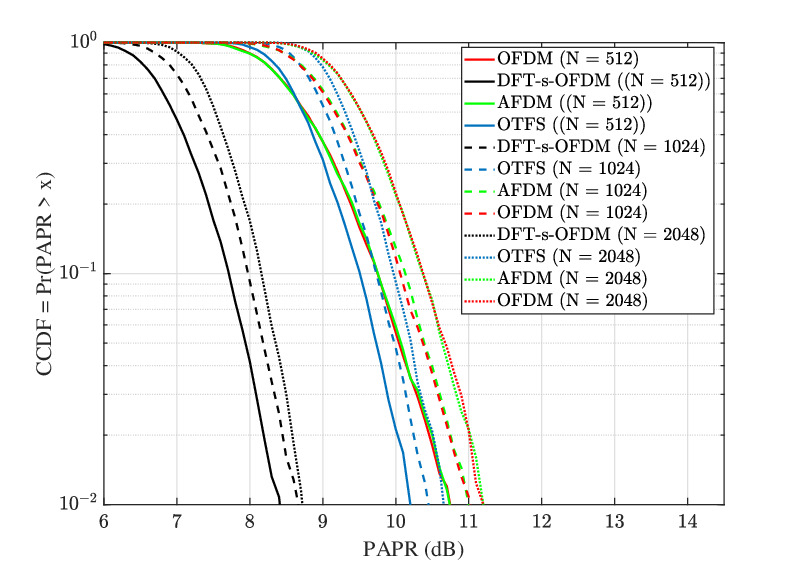}
    \caption{PAPR CCDF as function of different FFT sizes $N=\{512,1024,2048\}$.}
    \label{fig:PAPR_CCDF}
\end{figure}
Fig.~\ref{fig:PAPR_CCDF} shows the \ac{PAPR} CCDF for the considered waveforms and different transform sizes. As expected, \ac{DFT} s-\ac{OFDM} exhibits the lowest \ac{PAPR} across all transform sizes, since DFT spreading preserves a more single-carrier-like structure. \ac{OFDM} and \ac{AFDM} show comparable trends, while \ac{OTFS} yields a lower \ac{PAPR} than both of them over the considered settings. However, this comparison should be interpreted jointly with the required transform size. Under a fixed bandwidth, a fair operating comparison shows that \ac{OFDM} is typically stabilized by increasing $\Delta f$, which leads to shorter symbols and smaller required FFT sizes. In contrast, \ac{AFDM} and \ac{OTFS} generally require larger FFT sizes to achieve sufficient delay and Doppler resolution. Consequently, although the intrinsic \ac{PAPR} ordering remains favorable for \ac{DFT} s-\ac{OFDM} and moderate for \ac{OTFS}, the burden of \ac{AFDM} and \ac{OTFS} increases as they operate at larger transform sizes.\\
\begin{figure*}[t]
    \centering
    \begin{subfigure}[b]{0.20\textwidth}
        \centering
        \includegraphics[width=\textwidth]{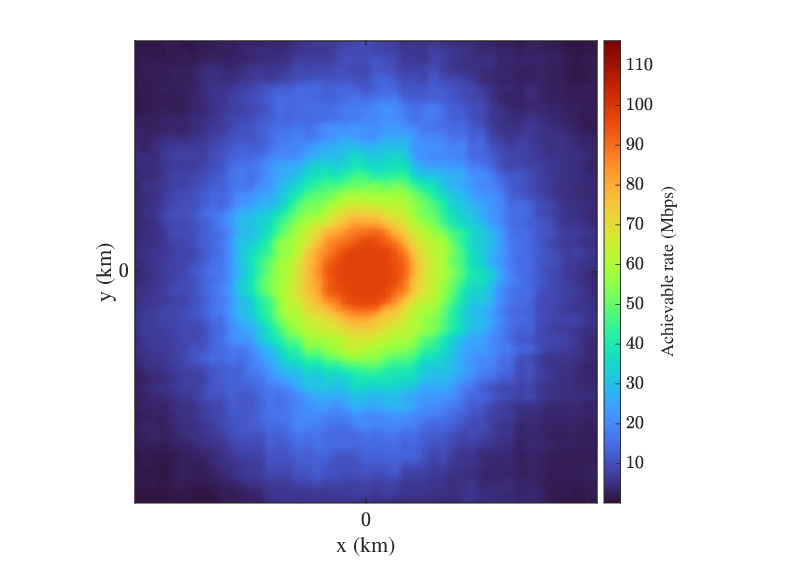}
        \caption{OFDM}
        \label{fig:AR_S_OFDM}
    \end{subfigure}\hfill
    \begin{subfigure}[b]{0.20\textwidth}
        \centering
        \includegraphics[width=\textwidth]{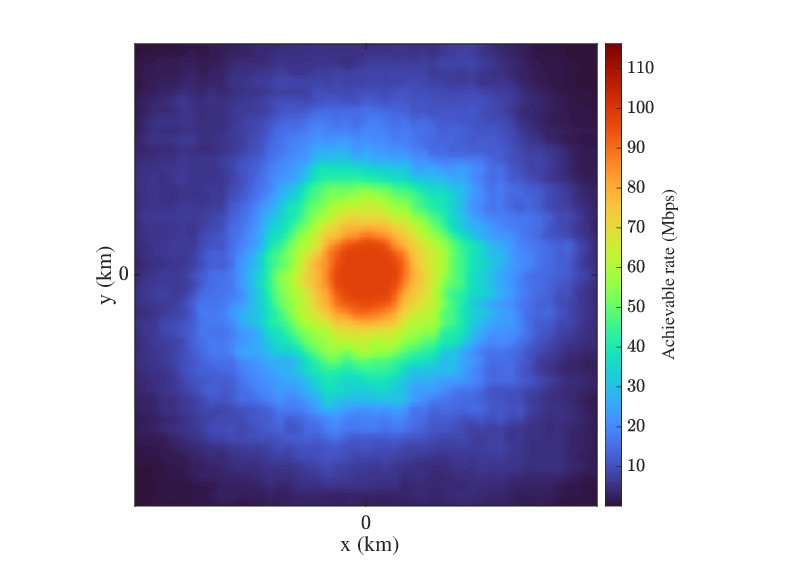}
        \caption{DFT s-OFDM}
        \label{fig:AR_S_DFT}
    \end{subfigure}\hfill
    \begin{subfigure}[b]{0.20\textwidth}
        \centering
        \includegraphics[width=\textwidth]{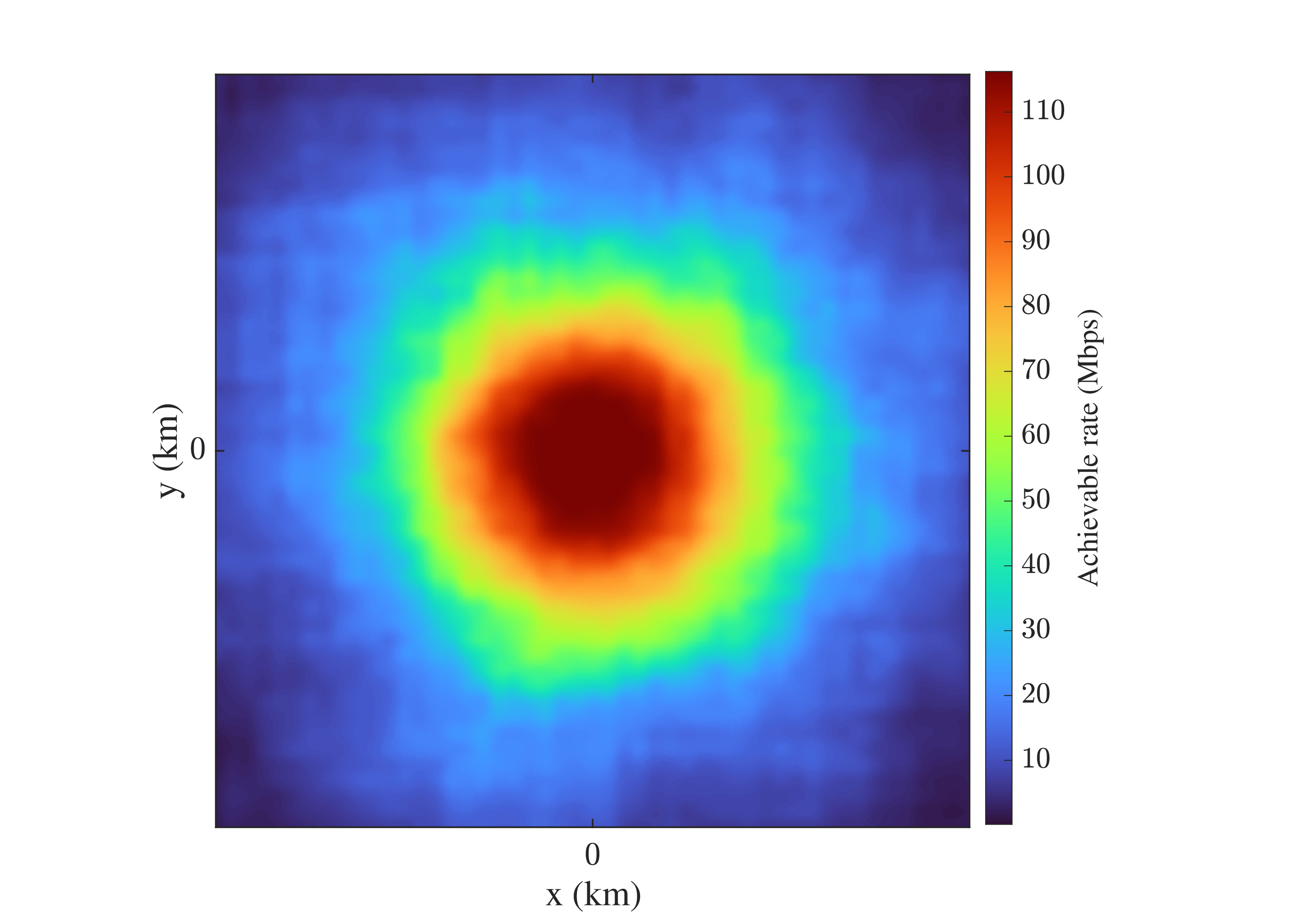}
        \caption{AFDM}
        \label{fig:AR_S_AFDM}
    \end{subfigure}\hfill
  \begin{subfigure}[b]{0.20\textwidth}
        \centering
        \includegraphics[width=\textwidth]{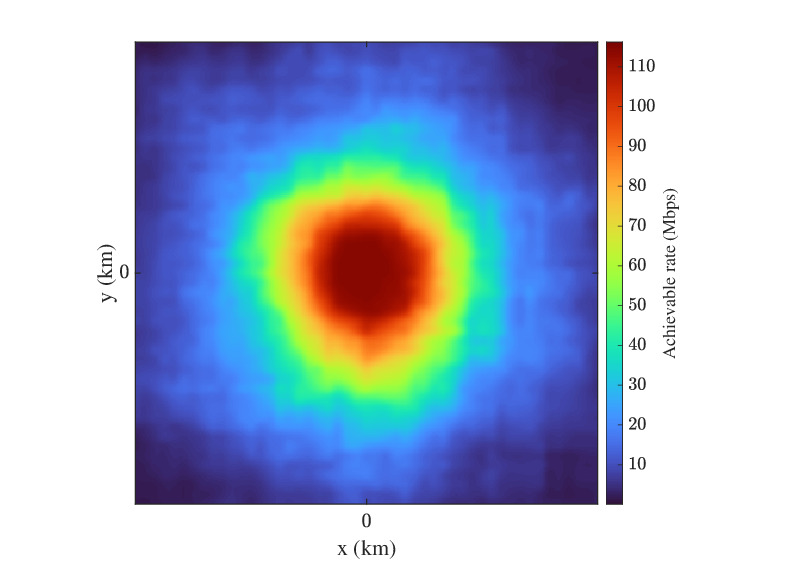}
        \caption{OTFS}
        \label{fig:AR_S_OTFS}
    \end{subfigure}
    \vspace{0.8em}

    \begin{subfigure}[b]{0.20\textwidth}
        \centering
        \includegraphics[width=\textwidth]{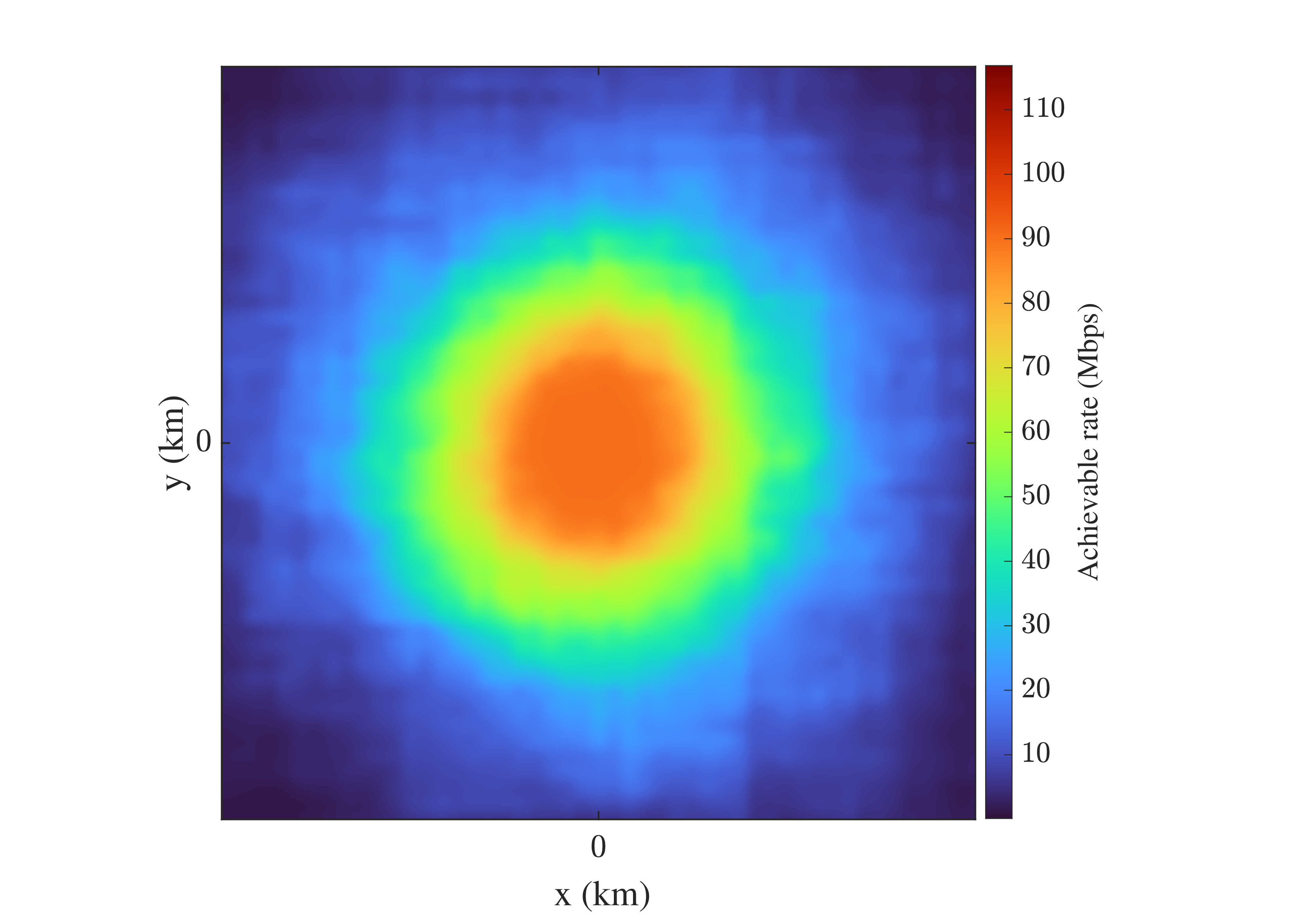}
        \caption{OFDM}
        \label{fig:AR_P_OFDM}
    \end{subfigure}\hfill
    \begin{subfigure}[b]{0.20\textwidth}
        \centering
        \includegraphics[width=\textwidth]{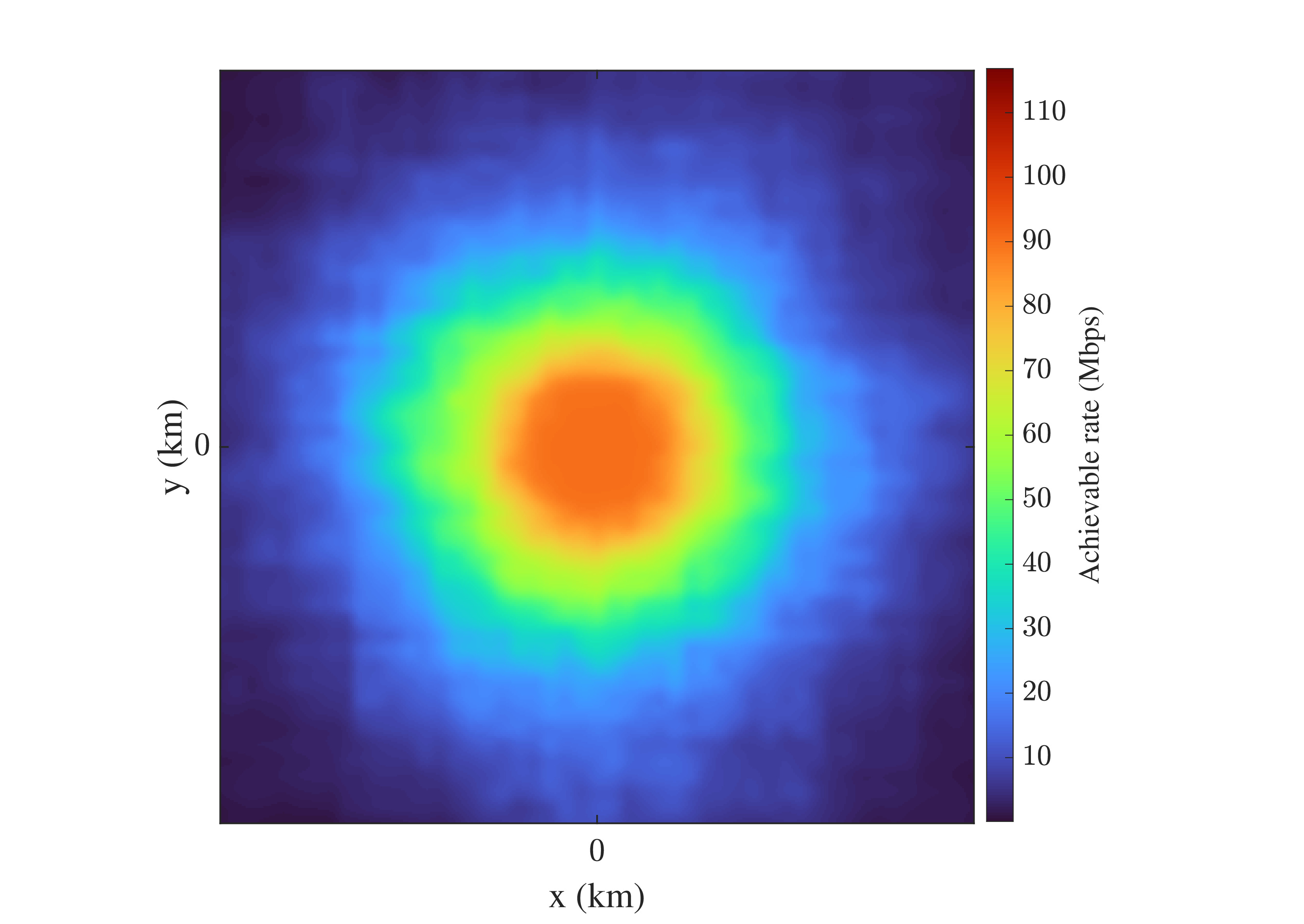}
        \caption{DFT s-OFDM}
        \label{fig:AR_P_DFT}
    \end{subfigure}\hfill
    \begin{subfigure}[b]{0.20\textwidth}
        \centering
        \includegraphics[width=\textwidth]{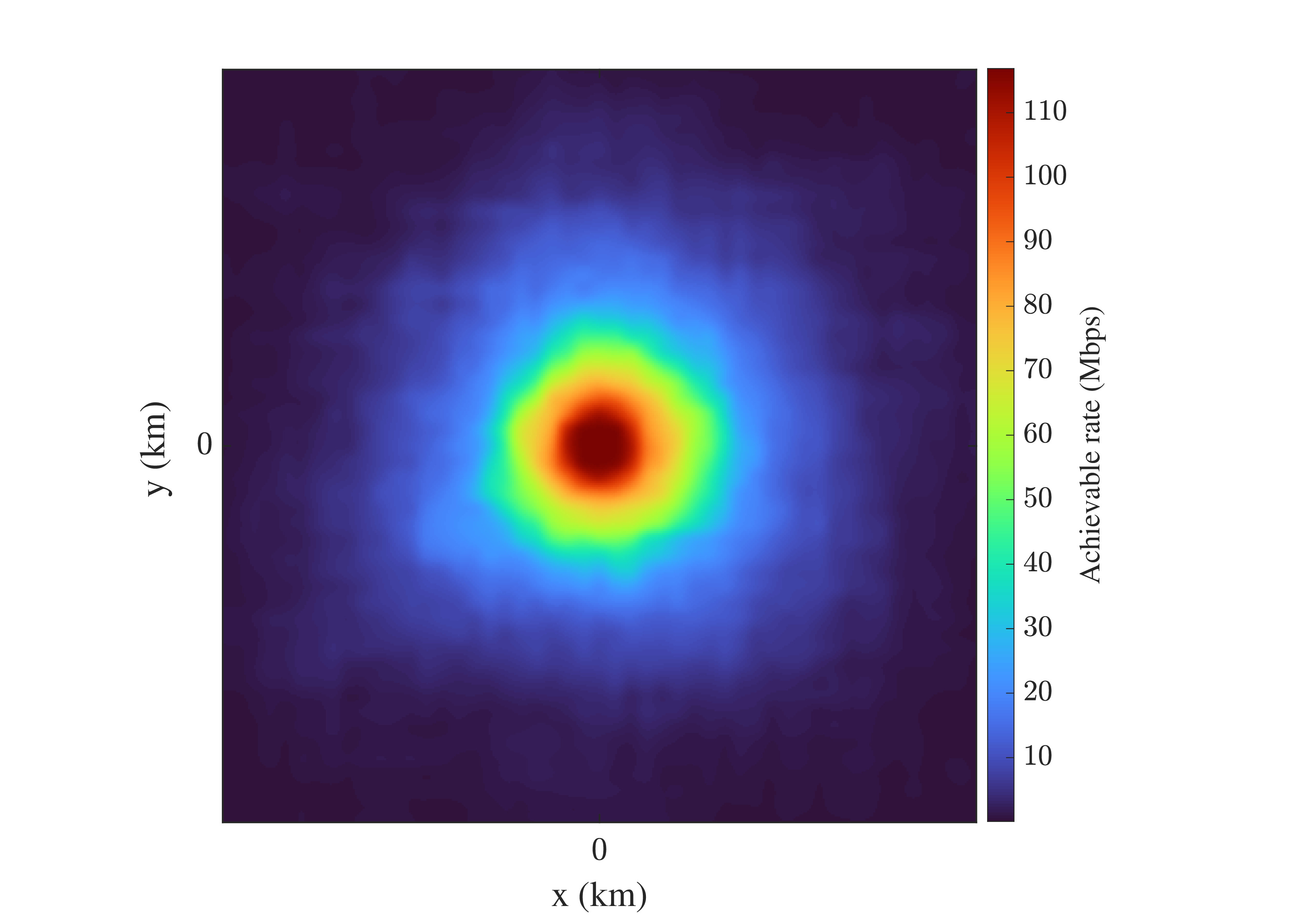}
        \caption{AFDM}
        \label{fig:AR_P_AFDM}
    \end{subfigure}\hfill
    \begin{subfigure}[b]{0.20\textwidth}
        \centering
        \includegraphics[width=\textwidth]{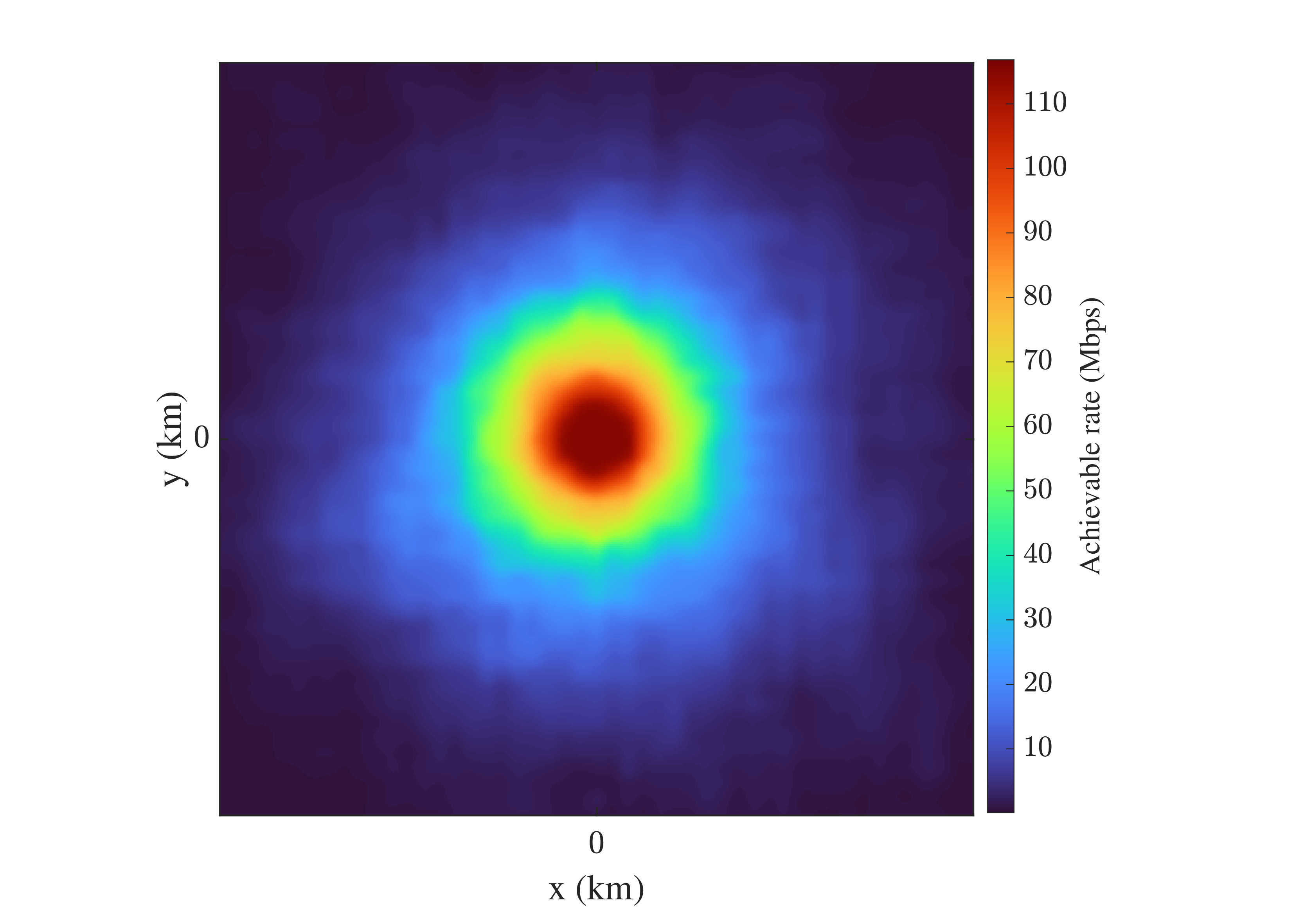}
        \caption{OTFS}
        \label{fig:AR_P_OTFS}
    \end{subfigure}
    \caption{Achievable-rate heat maps across the cell for the four waveforms: (a)-(d) sparse model and (e)-(h) proposed model.}
    \label{fig:Rate_map} 
\end{figure*}
Fig.~\ref{fig:Rate_map} presents the achievable-rate maps across the cell under both sparse and proposed channel models. The rate at position $\mathbf{r}$ is computed as $R(\mathbf{r})=B\,\eta\,\log_2\!\big(M^\star(\mathbf{r})\big)$, where $\eta$ is the waveform-dependent payload fraction. Therefore, the plotted rate depends jointly on propagation loss, waveform overhead, and the reliable modulation order that can be sustained at each cell location. Under the sparse model, where moderate or high useful \ac{SNR} can be preserved over a large region, \ac{AFDM} and \ac{OTFS} achieve the highest rates, especially near the cell center, since their sparse-domain processing enables more effective use of resolvable multipath structure. In contrast, \ac{OFDM} and \ac{DFT} s-\ac{OFDM} remain at lower achievable-rate levels, although their spatial distribution is still relatively stable. When moving to the proposed model, \ac{OFDM} and \ac{DFT} s-\ac{OFDM} retain nearly the same rate distribution over the cell, whereas \ac{AFDM} and \ac{OTFS} undergo a pronounced contraction of their high-rate regions. This again confirms that sparse-domain waveforms are advantageous only when the channel remains sufficiently resolvable and locally stationary, while \ac{OFDM}-like processing is more robust once leakage, segmentation, and support ambiguity dominate.

\section{Conclusion}
This paper puts together a generalized and mathematically traceable channel model that captures cluster birth-death phenomena, Doppler spectral spreading, time-evolving delays, and piecewise stationarity, providing a tractable bridge between idealized sparse representations and realistic environment propagation. Effective channel matrices were derived for OFDM, DFT-s-OFDM, AFDM, and OTFS, showing that waveform superiority is regime-dependent rather than universal. AFDM and OTFS exploit delay and Doppler structure only when sparsity and local stationarity hold, whereas OFDM and DFT-s-OFDM absorb channel effects more robustly through subcarrier spacing adaptation. These findings motivate channel-aware waveform selection criteria for 6G rather than a "one waveform for all", with future work targeting MIMO systems, ISAC channels, and hardware impairments.

\section{Appendix}
\appendices
\begin{figure*}
\normalsize
\begin{equation}\small \tag{76}
\begin{aligned}
 H^{\mathrm{DFT}}_{\mathrm{eff}}[m,n]
=\sum_{i,c,r}\frac{h_{c,r}^{(i)}}{N} 
\sum_{a=0}^{N_d-1}\sum_{b=0}^{N_d-1}F_{N_d}[m,a]
\sum_{p\in\mathcal{N}_i}\sum_{q=0}^{N-1}F[k_0+a,p]F^{\ast}[k_0+b,q]
\exp\!\left(j2\pi\frac{k_{c,r}^{\mathrm{tot}}}{N}p\right) 
\sum_{u=0}^{N-1}&\exp\!\left(j2\pi\frac{u}{N}(p-q-l_c^{\mathrm{tot},i})\right) \\ & \times
F_{N_d}^{\ast}[b,n].
\end{aligned}
\label{eq:app_sc_step2}
\end{equation}
\begin{equation}\small \tag{77}
\begin{aligned}
H^{\mathrm{DFT}}_{\mathrm{eff}}[m,n]
=\sum_{i,c,r}\frac{h_{c,r}^{(i)}}{N}
\sum_{u=0}^{N-1}\exp(-j2\pi\frac{u}{N}l_c^{tot,i}) 
\sum_{p\in\mathcal{N}_i}\sum_{q=0}^{N-1}
\exp(j2\pi\frac{k_{c,r}^{tot}+u}{N}p)
&\exp\!\left(-j2\pi\frac{u}{N}q\right) 
\sum_{a=0}^{N_d-1}F_{N_d}[m,a]\;F[k_0+a,p] 
\\ & \times\sum_{b=0}^{N_d-1}  F^{\ast}[q,k_0+b] \times F_{N_d}^{\ast}[b,n].
\end{aligned}
\label{eq:app_sc_step3}
\end{equation}
\hrulefill
\end{figure*}
\subsection*{A) General Effective-Channel Expansion }
For any transform $\mathbf{A}$, the effective channel is defined as
\begin{equation}
\mathbf{H}_{\mathrm{eff}}\triangleq \mathbf{A}\mathbf{H}\mathbf{A}^{H}.
\label{eq:app_Heff_def}
\end{equation}
Hence, its $(m,n)$th entry is
\begin{equation}
H_{\mathrm{eff}}[m,n]
=
\sum_{p=0}^{N-1}\sum_{q=0}^{N-1}
A[m,p]\;H[p,q]\;A^{\ast}[n,q].
\label{eq:app_Heff_entry}
\end{equation}
Substituting the piecewise structure and using \eqref{eq:DP_entry_expand} yields
\begin{equation} \small
\begin{aligned}
& H_{\mathrm{eff}}[m,n]
=\sum_{i=1}^{\mathcal{I}}\sum_{c=1}^{C_i}\sum_{r=1}^{R_{c,i}}
\frac{h_{c,r}^{(i)}}{N}
\sum_{p\in\mathcal{N}_i}\sum_{q=0}^{N-1}
A[m,p]\;A^{\ast}[n,q]\; \\ &
\exp\!\left(j2\pi\frac{k_{c,r}^{\mathrm{tot}}}{N}p\right)
\sum_{u=0}^{N-1}\exp\!\left(j2\pi\frac{u}{N}\big(p-q-l_{c}^{\mathrm{tot},i}\big)\right).
\end{aligned}
\label{eq:app_Heff_mid1}
\end{equation}
The phase term is factored as
\begin{equation} \small
\begin{aligned}
   & \exp\!\left(j2\pi\frac{u}{N}\big(p-q-l_{c}^{\mathrm{tot},i}\big)\right)
= 
\exp\!\left(j2\pi\frac{u}{N}p\right)\;\\ & \times
\exp\!\left(-j2\pi\frac{u}{N}q\right)\;  \times
\exp\!\left(-j2\pi\frac{u}{N}l_{c}^{\mathrm{tot},i}\right),
\label{eq:app_phase_factor}
\end{aligned}
\end{equation}
and the order of summations is exchanged, which gives
\begin{equation} \small
\begin{aligned}
H_{\mathrm{eff}}[m,n]
&=\sum_{i=1}^{\mathcal{I}}\sum_{c=1}^{C_i}\sum_{r=1}^{R_{c,i}}
\frac{h_{c,r}^{(i)}}{N}
\sum_{u=0}^{N-1}\exp\!\left(-j2\pi\frac{u}{N}l_{c}^{\mathrm{tot},i}\right)\\
&\quad\times
\sum_{p\in\mathcal{N}_i}A[m,p]\;\exp\!\left(j2\pi\frac{k_{c,r}^{\mathrm{tot}}+u}{N}p\right) \\ & \times 
\sum_{q=0}^{N-1}A^{\ast}[n,q]\;\exp\!\left(-j2\pi\frac{u}{N}q\right).
\end{aligned}
\label{eq:app_Heff_mid2}
\end{equation}
Substituting the corresponding kernel entries $A[\cdot,\cdot]$ into \eqref{eq:app_Heff_mid2} yields the waveform specific expressions reported in Sec.~III.
\subsection*{B) Entry Wise Expansion of Fractional Delay Operator}
Starting from the definition of $\mathbf{P}(l)$ given in the I/O section, the $(p,q)$ entry is
\begin{equation}
\begin{aligned}
P(l)[p,q]
&=\sum_{m=0}^{N-1}F^{\ast}[m,p]\;\Lambda(l)[m,m]\;F[m,q] \\
&=\frac{1}{N}\sum_{m=0}^{N-1}e^{j2\pi \frac{m}{N}p}\;e^{-j2\pi \frac{m}{N}l}\;e^{-j2\pi \frac{m}{N}q} \\
&
=\frac{1}{N}\sum_{u=0}^{N-1}\exp\!\left(j2\pi\frac{u}{N}(p-q-l)\right).
\end{aligned}
\label{eq:app_P_entry}
\end{equation}
Here, $u$ is the auxiliary DFT-domain summation index that arises from the factorization of the fractional-delay operator $\mathbf{P}(l)=\mathbf{F}^{H}\mathbf{\Lambda}(l)\mathbf{F}$.
\subsection*{C) OTFS Index Mapping Step}
Using the index mapping $p=a+l M'$ and $q=b+l' M'$ with $M'N'=N$, the OTFS kernel entries satisfy
\begin{equation}
\begin{aligned}
    A_{\mathrm{otfs}}\!\big[(\mu,a),a+l M'\big]=\frac{1}{\sqrt{N'}}e^{-j2\pi\frac{\mu l}{N'}},
\\ \qquad
A_{\mathrm{otfs}}^{\ast}\!\big[(\nu,b),b+l' M'\big]=\frac{1}{\sqrt{N'}}e^{+j2\pi\frac{\nu l'}{N'}},
\end{aligned}
\label{eq:app_Aotfs_entries}
\end{equation}
and the element-wise similarity expansion becomes
\begin{equation}
\begin{aligned}
    H^{\mathrm{dd}}_{\mathrm{eff}}\!\big[(\mu,a),(\nu,b)\big]
& =\frac{1}{N'}\sum_{l=0}^{N'-1}\sum_{l'=0}^{N'-1}
e^{-j2\pi\frac{\mu l}{N'}}\;e^{+j2\pi\frac{\nu l'}{N'}}\; \\ &
H\!\big[a+l M',\,b+l' M'\big],
\end{aligned}
\label{eq:app_Heff_otfs_mid1}
\end{equation}
which is the starting point used in Sec.~III before substituting the scalable channel entry.
\subsection*{D) Steps For DFT s-OFDM Effective Channel}
Starting from \eqref{eq:Heff_sc_mid0} and using the element-wise representation of $\mathbf{F}\mathbf{H}\mathbf{F}^{H}$,
\begin{equation}
\begin{aligned}
 H^{\mathrm{DFT}}_{\mathrm{eff}}[m,n]
=\sum_{a=0}^{N_d-1}\sum_{b=0}^{N_d-1}&F_{N_d}[m,a] 
\sum_{p=0}^{N-1}\sum_{q=0}^{N-1}F[k_0+a,p] \\ & \times H[p,q]\;F^{\ast}[q,k_0+b] F_{N_d}^{\ast}[b,n].   
\end{aligned}
\label{eq:app_sc_step1}
\end{equation}
Substituting the piecewise channel structure and using \eqref{eq:DP_entry_expand} yields to ($76$). Using \eqref{eq:app_phase_factor} and exchanging the order of summations yields to ($77$). At the end, substituting the DFT kernel entries and collecting phase terms yields \eqref{eq:Heff_sc_final}.
\bibliographystyle{IEEEtran}
\bibliography{Mendeley}


\end{document}